\documentclass[a4paper,12pt]{article}

\pdfoutput=1

\usepackage{amsmath}
\usepackage{amssymb}
\usepackage{amsfonts}
\usepackage{mathrsfs}
\usepackage{bbm}
\usepackage{graphicx,subfigure,booktabs,adjustbox}
\usepackage[numbers,sort&compress]{natbib}
\usepackage{verbatim}

\usepackage{color}
\usepackage{ulem}
\usepackage{setspace}
\usepackage{multirow}

\usepackage[colorlinks,
            linkcolor=black,
            filecolor=black,
            anchorcolor=black,
            urlcolor=black,
            citecolor=black,
            bookmarks=false,
            ]{hyperref}

\numberwithin{equation}{section}

\newlength{\dinwidth}
\newlength{\dinmargin}
\makeatletter
\newcommand{\thickhline}{%
    \noalign {\ifnum 0=`}\fi \hrule height 1pt
    \futurelet \reserved@a \@xhline
}
\makeatother

\allowdisplaybreaks

\begin{document}

\title{\bf \Large \boldmath{$\Lambda_b\to\Lambda_c\tau\bar\nu_\tau$} decay in scalar and vector leptoquark scenarios}

\author{
Xin-Qiang Li\footnote{xqli@mail.ccnu.edu.cn},
Ya-Dong Yang\footnote{yangyd@mail.ccnu.edu.cn}\,
and
Xin Zhang\footnote{zhangxin027@mails.ccnu.edu.cn}\\[15pt]
\small Institute of Particle Physics and Key Laboratory of Quark and Lepton Physics~(MOE), \\
\small Central China Normal University, Wuhan, Hubei 430079, China}

\date{}
\maketitle
\vspace{0.2cm}

\begin{abstract}
{\noindent}It has been shown that the anomalies observed in $\bar B\to D^{(\ast)}\tau\bar\nu_\tau$ and $\bar B\to \bar K\ell^+\ell^-$ decays can be resolved by adding a single scalar or vector leptoquark to the Standard Model, while constraints from other precision measurements in the flavour sector can be satisfied without fine-tuning. To further explore these two interesting scenarios, in this paper, we study their effects in the semi-leptonic $\Lambda_b\to\Lambda_c\tau\bar\nu_\tau$ decay. Using the best-fit solutions for the operator coefficients allowed by the current data of mesonic decays, we find that (i) the two scenarios give similar amounts of enhancements to the branching fraction $\mathcal B(\Lambda_b\to\Lambda_c\tau\bar\nu_\tau)$ and the ratio $R_{\Lambda_c}=\mathcal B(\Lambda_b\to\Lambda_c \tau\bar\nu_\tau)/\mathcal B(\Lambda_b\to\Lambda_c\ell\bar\nu_\ell)$, (ii) the two best-fit solutions in each of these two scenarios are also indistinguishable from each other, (iii) both scenarios give nearly the same predictions as those of the Standard Model for the longitudinal polarizations of $\Lambda_c$ and $\tau$ as well as the lepton-side forward-backward asymmetry. With future measurements of these observables in $\Lambda_b\to\Lambda_c\tau\bar\nu_\tau$ decay at the LHCb, the two leptoquark scenarios could be further tested, and even differentiated from the other NP explanations for the $R_{D^{(\ast)}}$ anomalies. We also discuss the feasibility for the measurements of these observables at the LHC and the future $e^+e^-$ colliders.
\end{abstract}

\newpage

\section{Introduction}
\label{sec:intro}

While no direct evidences for physics beyond the Standard Model (SM) have been found at the LHC so far, there are some interesting indirect hints for New Physics (NP) in the flavour sector~\cite{Crivellin:2016ivz,Ligeti:2016riq,Ricciardi:2016pmh}. It is particularly interesting to note that intriguing effects of lepton favour universality violation (LFUV) have been observed in rare $B$-meson decays. To be more specific, the ratios of charged-current decays, $R_{D}=\frac{{\mathcal B}(\bar B\to D\tau\bar\nu_\tau)}
{{\mathcal B}(\bar B\to D\ell\bar\nu_\ell)}$ and $R_{D^\ast}=\frac{{\mathcal B}(\bar B\to D^\ast\tau\bar\nu_\tau)}
{{\mathcal B}(\bar B\to D^\ast\ell\bar\nu_\ell)}$, with $\ell=e,\mu$, have been measured by the BaBar~\cite{Lees:2012xj,Lees:2013uzd}, Belle~\cite{Huschle:2015rga,Abdesselam:2016cgx,Abdesselam:2016xqt} and LHCb~\cite{Aaij:2015yra} Collaborations. The latest averages by the Heavy Flavor Average Groups (HFAG)~\cite{HFAG:2016},
\begin{equation}
R_D^{\mathrm{exp.}}=0.397\pm0.040({\mathrm{stat.}})\pm0.028({\mathrm{syst.}})\,, \quad
R_{D^{\ast}}^{\mathrm{exp.}}=0.316\pm0.016({\mathrm{stat.}})\pm0.010({\mathrm{syst.}})\,,
\end{equation}
exceed the SM predictions,
\begin{align}
R_D^{\mathrm{SM}}=0.300 \pm 0.008~\text{\cite{Na:2015kha}}\,, \qquad R_{D^{\ast}}^{\mathrm{SM}}=0.252 \pm 0.003~\text{\cite{Fajfer:2012vx}}\,,
\end{align}
by $1.9\sigma$ and $3.3\sigma$, respectively. Once the measurement correlations between $R_D$ and $R_{D^\ast}$ are taken into account, the deviation will be at $4.0\sigma$ level. Another hint of LFUV has also been reported in the $b\to s \ell^+\ell^-$ process by the LHCb experiment~\cite{Aaij:2014ora}:
\begin{equation}
R_K^{\mathrm{exp.}}=\frac{{\Gamma}(B^+\to K^+\mu^+\mu^-)}
{{\Gamma}(B^+\to K^+ e^+e^-)}\bigg|_{q^2\in [1,6]\,{\rm GeV}^2}=0.745_{-0.074}^{+0.090}({\mathrm{stat.}})\pm0.036({\mathrm{syst.}})\,,
\end{equation}
which is about $2.6\sigma$ lower than the corresponding SM prediction $R_K^{\rm SM}=1.00\pm0.03$~\cite{Hiller:2003js,Bordone:2016gaq}.

The observed $R_{D^{(\ast)}}$ and $R_K$ anomalies, if confirmed with future more precise data, would be clear signs for NP beyond the SM, and have already inspired lots of studies; for a recent review, the readers are referred to Refs.~\cite{Crivellin:2016ivz,Ligeti:2016riq,Ricciardi:2016pmh} and references therein. Here we are interested in the possible NP solutions with a single scalar or vector leptoquark (LQ) scenario~\cite{Bauer:2015knc,Fajfer:2015ycq}. In Ref.~\cite{Bauer:2015knc}, it has been shown that the anomalies $R_{D^{(\ast)}}$, $R_K$ and $(g-2)_\mu$ could be addressed by adding to the SM just one TeV-scale scalar LQ transforming as $(\mathbf{3}, \mathbf{1},-\frac{1}{3})$ under the SM gauge group. On the other hand, as shown in Ref.~\cite{Fajfer:2015ycq}, the $R_{D^{(\ast)}}$, $R_K$ and the angular observable $P_5^\prime$ in $\bar B\to\bar K^\ast\mu^+\mu^-$ decay could be explained by just one vector LQ transforming as $(\mathbf{3}, \mathbf{3},\frac{2}{3})$ under the SM gauge group. Under the constraints from both the ratios $R_{D^{(\ast)}}$ and the $q^2$ spectra of $\bar B\to D^{(\ast)}\tau\bar\nu_\tau$ decays provided by the BaBar~\cite{Lees:2013uzd} and Belle~\cite{Huschle:2015rga,Abdesselam:2016cgx} Collaborations, four best-fit solutions are found for the operator coefficients induced by the scalar LQ~\cite{Freytsis:2015qca}, two of which are, however, already excluded by the purely leptonic $B_c^-\to \tau^-\bar\nu_\tau$ decay~\cite{Li:2016vvp}. At the same time, two best-fit solutions are also found for the operator coefficients induced by the vector LQ~\cite{Freytsis:2015qca}.

To further explore the two interesting LQ scenarios, in this paper, we shall study their effects in the semi-leptonic $\Lambda_b\to\Lambda_c\tau\bar\nu_\tau$ decay, which is induced by the same quark-level transition as the $\bar B\to D^{(\ast)}\tau\bar\nu_\tau$ decays. While the $\Lambda_b$ baryons are not produced at an $e^+ e^-$ $B$-factory, they account for around $20\%$ of the $b$-hadrons produced at the LHC~\cite{Aaij:2011jp}. Remarkably, the produced number of $\Lambda_b$ baryons is comparable to that of $B_u$ or $B_d$ mesons, and is significantly higher than that of $B_s$ meson~\cite{Aaij:2011jp,Meinel:2014wua}. Due to the spin-half nature of $\Lambda_b$, its decay may provide complementary information compared to the corresponding mesonic one. Motivated by the $R_{D^{(\ast)}}$ anomalies, the semi-leptonic $\Lambda_b\to\Lambda_c\ell\bar\nu_\ell$ decay has been studies recently in Refs.~\cite{Woloshyn:2014hka,Shivashankara:2015cta,Gutsche:2015mxa,Dutta:2015ueb,Detmold:2015aaa,DiSalvo:2016rfi}.

In this paper, besides the total and differential branching fractions, as well as the ratio $R_{\Lambda_c}=\frac{{\mathcal B}(\Lambda_b\to\Lambda_c\tau\bar\nu_\tau)}{{\mathcal B}(\Lambda_b\to\Lambda_c\ell\bar\nu_\ell)}$, discussed already in previous studies~\cite{Shivashankara:2015cta,Gutsche:2015mxa,Dutta:2015ueb,Detmold:2015aaa}, we shall also discuss the longitudinal polarizations of the daughter baryon $\Lambda_c$ and the $\tau$ lepton, and the lepton-side forward-backward asymmetry in this decay~\cite{Gutsche:2015mxa}. The feasibility for the measurements of these observables at the LHC, which is the currently available experiment to explore the $\Lambda_b$ decays, as well as at the future $e^+e^-$ colliders, such as the International Linear Collider (ILC) and the Circular Electron Positron Collider (CEPC), will also be discussed. We calculate these observables using the helicity formalism developed in Refs.~\cite{Korner:1989v,Korner:1989qb}, and have rederived and confirmed the helicity amplitudes associated with both the (axial-)vector and (pseudo-)scalar interactions given already in Refs.~\cite{Shivashankara:2015cta,Gutsche:2015mxa,Dutta:2015ueb}; for the  (pseudo-)tensor-type current, however, the corresponding helicity amplitudes are new and presented here for the first time. As the $\Lambda_b\to\Lambda_c$ transition form factors are not yet determined quite well and still bring large uncertainties, it is instructive to check the sensitivity of these observables to the different values of form factors obtained, for example, in a covariant confined quark model~\cite{Gutsche:2015mxa} (used in Ref.~\cite{Gutsche:2015mxa}), in the QCD sum rules~\cite{MarquesdeCarvalho:1999bqs} (used in Ref.~\cite{Shivashankara:2015cta}), or in the lattice calculations~\cite{Detmold:2015aaa} (used in Refs.~\cite{Dutta:2015ueb,Detmold:2015aaa}). To this end, rather than choosing a single form for these form factors, we use, as a comparison, the results obtained both from the QCD sum rules~\cite{MarquesdeCarvalho:1999bqs}, which satisfy the heavy quark effective theory (HQET) relations~\cite{Neubert:1993mb,Falk:1991nq,Mannel:1990vg}, and from the latest lattice calculations with $2+1$ dynamical flavours~\cite{Detmold:2015aaa}\footnote{There are currently only the lattice results for the (axial-)vector form factors~\cite{Detmold:2015aaa}. For the (pseudo-)tensor form factors, since there are no lattice results yet, we still use the HQET relations to relate them to the corresponding (axial-)vector ones.}. Using the best-fit solutions for the operator coefficients allowed by the current data of mesonic decays, we find that the two scenarios give similar amounts of enhancements relative to the SM predictions for the branching fraction $\mathcal B(\Lambda_b\to\Lambda_c\tau\bar\nu_\tau)$ and the ratio $R_{\Lambda_c}$, and the two best-fit solutions in each of the two scenarios are also indistinguishable from each other based only on these two observables. On the other hand, both of these two scenarios give nearly the same predictions as the SM for the longitudinal polarizations of $\Lambda_c$ and $\tau$ as well as the lepton forward-backward asymmetry. With future precise measurements of these observables at the LHCb, the two scenarios could be further tested and even differentiated from the other explanations to the $R_{D^{(\ast)}}$ anomalies~\cite{Hou:1992sy,Tanaka:1994ay,Kiers:1997zt,Chen:2005gr,Nierste:2008qe,Faller:2011nj,Bailey:2012jg,
Becirevic:2012jf,Datta:2012qk,Celis:2012dk,Tanaka:2012nw,Fajfer:2012jt,Deshpande:2012rr,Sakaki:2012ft,
Crivellin:2013wna,Sakaki:2013bfa,Dutta:2013qaa,Duraisamy:2013kcw,Biancofiore:2013ki,Fan:2013qz,Bhattacharya:2014wla,
Duraisamy:2014sna,Hagiwara:2014tsa,Sakaki:2014sea,Sahoo:2015pzk,Barbieri:2015yvd,Bhattacharya:2015ida,
Freytsis:2015qca,Calibbi:2015kma,Cline:2015lqp,Kim:2015zla,Crivellin:2015hha,Hwang:2015ica,Alonso:2015sja,
Hati:2015awg,Greljo:2015mma,Fan:2015kna,Becirevic:2016hea,Alok:2016qyh,Ivanov:2016qtw,Deppisch:2016qqd,
Dumont:2016xpj,Boucenna:2016wpr,Das:2016vkr,Zhu:2016xdg,Wang:2016ggf,Kim:2016yth,Bardhan:2016uhr,
Ligeti:2016npd,Hiller:2016kry,Faroughy:2016osc,Sahoo:2016pet,Becirevic:2016yqi,Deshpand:2016cpw,Boucenna:2016qad,
Bhattacharya:2016mcc,Dutta:2016em,Bhattacharya:2016zcw,Alonso:2016oyd,Choudhury:2016ulr,Celis:2016azn,Ivanov:2017mrj}.

This paper is organized as follows: In section~\ref{sec:framework}, we recapitulate briefly both the scalar and vector LQ scenarios~\cite{Bauer:2015knc,Fajfer:2015ycq}. In section~\ref{sec:analytic calculation}, we calculate the helicity amplitudes and list the relevant physical observables for the semi-leptonic $\Lambda_b\to\Lambda_c\ell\bar\nu_\ell$ decays. In section~\ref{sec:numerical results}, the scalar and vector LQ effects on the branching fraction $\mathcal B(\Lambda_b\to\Lambda_c\tau\bar\nu_\tau)$, the ratio $R_{\Lambda_c}$, the $\Lambda_c$ and $\tau$ longitudinal polarizations, as well as the lepton-side forward-backward asymmetry are discussed. We finally conclude in section~\ref{sec:conclusions}. The $\Lambda_b\to\Lambda_c$ transition form factors and the helicity-dependent differential decay rates are collected in Appendices~\ref{sec:form factors} and \ref{sec:helicity rates}, respectively.

\section{The scalar and vector LQ scenarios}
\label{sec:framework}

In this section, we recapitulate the scalar and vector LQ models, where a single TeV-scale scalar or vector LQ is added to the SM to address the aforementioned anomalies~\cite{Bauer:2015knc,Fajfer:2015ycq}. For a recent comprehensive review of LQ models, the readers are referred to Ref.~\cite{Dorsner:2016wpm}.

\subsection{The scalar LQ scenario}\label{subsec:scalar}

Firstly, we consider the scalar LQ $\phi$ transforming as $(\mathbf{3}, \mathbf{1}, -\frac{1}{3})$ under the SM gauge group, in which its couplings to SM fermions are described by the Lagrangian~\cite{Bauer:2015knc}
\begin{equation}\label{eq:scalar coupling}
\mathcal L_{\rm int}^\phi=\bar Q_L^c{\boldsymbol\lambda}^Li\tau_2L\phi^{\ast}+\bar u_R^c{\boldsymbol\lambda}^R\ell_R\phi^{\ast}+{\rm h.c.}\,,
\end{equation}
where ${\boldsymbol\lambda}^{L,R}$ are the Yukawa coupling matrices in flavour space, and $Q_L,\,L$ denote the left-handed quark and lepton doublet, while $u_R,\,\ell_R$ the right-handed up-type quark and lepton singlet, respectively. The charge-conjugated spinors are defined as $\psi^c=C\bar{\psi}^T$, $\bar{\psi}^c=\psi^T C$~($C=i\gamma^2\gamma^0$). Such a scalar $\phi$ mediates the $b\to c\tau\bar\nu_\tau$ decay at tree level, and the resulting effective weak Hamiltonian including the SM contribution is given as~\cite{Bauer:2015knc,Li:2016vvp}
\begin{align}\label{eq:scalar Hamiltonian}
\mathcal H_{\text{eff}}=&\frac{4G_FV_{cb}}{\sqrt2}\left[C_V\,\bar c\gamma_\mu P_L b\,\bar\tau\gamma^\mu P_L\nu_\tau+C_S\,\bar c P_L b\,\bar\tau P_L\nu_\tau-\frac{1}{4}C_T\,
\bar c\sigma_{\mu\nu}P_Lb\,\bar{\tau}\sigma^{\mu\nu}P_L\nu_{\tau}\right]\,,
\end{align}
where $C_V$, $C_S$, $C_T$ are the Wilson coefficients of the corresponding four-fermion operators and, at the matching scale $\mu=M_\phi$, are given explicitly as
\begin{align}
&C_V(M_\phi)=1+\frac{\lambda_{b\nu_{\tau}}^L\lambda_{c\tau}^{L\ast}}
{4\sqrt{2}G_FV_{cb}M_{\phi}^2}\,,\\[0.2cm]
&C_S(M_\phi)=C_T(M_\phi)=-\frac{\lambda_{b\nu_{\tau}}^L\lambda_{c\tau}^{R\ast}}
{4\sqrt{2}G_FV_{cb}M_{\phi}^2}\,.\label{eq:WCmuphi}
\end{align}
In order to resum potentially large logarithmic effects, the Wilson coefficients $C_S$ and $C_T$ should be run down to the characteristic scale of the process we are interested in, {\it i.e.}, $\mu_b\sim m_b$, while $C_V$ is not renormalized because of the conservation of vector currents. The explicit evolution equations could be found, for example, in Ref.~\cite{Dorsner:2016wpm}.

As shown in Ref.~\cite{Bauer:2015knc}, such a scalar LQ could explain the $R_{D^{(\ast)}}$, $R_K$ and $(g-2)_\mu$ anomalies, while constraints from other precision measurements in the flavour sector can be satisfied without fine-tuning. Especially, under the constraints from both the ratios $R_{D^{(\ast)}}$ and the measured $q^2$ spectra in $B\to D^{(\ast)}\tau\bar\nu_{\tau}$ decays, four best-fit solutions are found for the operator coefficients induced by the scalar LQ~\cite{Freytsis:2015qca}, two of which are, however, already excluded by the purely leptonic $B_c^-\to \tau^-\bar\nu_\tau$ decay~\cite{Li:2016vvp}. Consequently, in this paper, we shall consider only the remaining two solutions denoted by $P_A$ and $P_C$ in Ref.~\cite{Li:2016vvp}.

\subsection{The vector LQ scenario}

We now introduce the second scenario in which the SM is extended by a vector $SU(2)_L$ triplet $U_3^\mu$ transforming as $(\mathbf{3}, \mathbf{3},\frac{2}{3})$ under the SM gauge group. The coupling of the vector multiplet $U_3^\mu$ to a lepton-to-quark current with $(V-A)$ structure is given by~\cite{Fajfer:2015ycq}
\begin{equation}\label{eq:vector coupling}
\mathcal L_{U_3}=g_{ij}\bar Q_i\gamma^\mu\tau^A U^A_{3\mu}L_j+{\rm h.c.}\,,
\end{equation}
where $\tau^A$~($A=1,2,3$) are the Pauli matrices in the $SU(2)_L$ space, and $L_i$ and $Q_i$~($i,j=1,2,3$) \par \noindent denote the left-handed lepton and quark doublets, respectively. The Lagrangian Eq.~\eqref{eq:vector coupling} is written in the fermion mass basis, with $g_{ij}$ defined as the couplings of the $Q=2/3$ component of the triplet, $U_{3\mu}^{(2/3)}$, to $\bar d_{Li}$ and $\ell_{Lj}$. Expanding the $SU(2)_L$ components, we get explicitly
 \begin{align}
 \mathcal L_{U_3} &= U_{3\mu}^{(2/3)}\big[\mathcal Vg\mathcal U)_{ij}\bar u_i\gamma^\mu P_L\nu_j-g_{ij}\bar d_i\gamma^\mu P_L\ell_j\big]\,\nonumber\\[0.2cm]
 &\quad +U_{3\mu}^{(5/3)}(\sqrt2\mathcal Vg)_{ij}\bar u_i\gamma^\mu P_L\ell_j\,\nonumber\\[0.2cm]
 &\quad +U_{3\mu}^{(-1/3)}(\sqrt2g\mathcal U)_{ij}\bar d_i\gamma^\mu P_L\nu_j+{\rm h.c.}\,,
 \end{align}
where $\mathcal V$ and $\mathcal U$ represent the Cabibbo-Kobayashi-Maskawa (CKM)~\cite{Cabibbo:1963yz,Kobayashi:1973fv} and the Pontecorvo-Maki-Nakagawa-Sakata (PMNS)~\cite{Pontecorvo:1957qd,Maki:1962mu} matrix, respectively. Here we assume the neutrinos to be massless and, therefore, the PMNS matrix can be rotated away through field redefinitions.

The vector multiplet $U_3^\mu$ can also mediate the $b\to c\tau\bar\nu_\tau$ transitions at tree level, and the resulting effective weak Hamiltonian including the SM contribution can be written as~\cite{Fajfer:2015ycq}
\begin{equation}\label{eq:vector Hamiltonian}
\mathcal H_{\rm eff}=\frac{4G_F V_{cb}}{\sqrt2}C_V^\prime(\bar c\gamma^\mu P_L b)(\bar\tau\gamma_\mu P_L\nu_\tau)\,,
\end{equation}
where $C_V^\prime$ is the Wilson coefficient at the matching scale $\mu=M_U$ and is given by
\begin{equation}\label{eq:vector WC}
C_V^\prime=1+\frac{\sqrt2g_{b\tau}^\ast(\mathcal Vg)_{c\tau}}{4G_FV_{cb}M_U^2}\,.
\end{equation}
Unlike in the scalar LQ case, the vector LQ only generates $(V-A)$ couplings and, therefore, the Wilson coefficient $C_V^\prime$ need not be renormalized.

As shown in Ref.~\cite{Fajfer:2015ycq}, the vector LQ scenario could also accommodate the $R_{D^{(\ast)}}$, $R_K$ as well as the angular observable $P_5^\prime$ in $\bar B\to \bar K^\ast\mu^+\mu^-$ decay. Fitting to the measured ratios $R_{D^{(\ast)}}$, along with acceptable $q^2$ spectra, two best-fit solutions, denoted as $R_A$ and $R_B$, respectively, are found in this scenario~\cite{Freytsis:2015qca}:
\begin{align}\label{eq:parameter}
g_{b\tau}^\ast(\mathcal Vg)_{c\tau} =&
\Bigg\{
      \begin{array}{rl}
        0.18\pm0.04,  & R_A\\
        -2.88\pm0.04,  & R_B
      \end{array}
    \Bigg.\,,
\end{align}
where $M_U= 1\,{\rm TeV}$ is taken as a benchmark. It should be noted that the triplet nature of $U_3^{\mu}$ also leads to various charged lepton-flavour-violating decays, such as the $B \to K\mu\tau$ and $\Upsilon(nS)\to\tau\mu$ decays, which have been discussed in Refs.~\cite{Fajfer:2015ycq,Bhattacharya:2016mcc}.

\section{$\Lambda_b\to\Lambda_c\ell\bar\nu_\ell$ decays in scalar and vector LQ scenarios}
\label{sec:analytic calculation}

\subsection{Helicity amplitudes}

In this subsection, we give the helicity amplitudes for the process $\Lambda_b\to\Lambda_c\ell\bar\nu_\ell$ both within the SM and in the two LQ scenarios. Following Refs.~\cite{Korner:1991ph,Gutsche:2015mxa} and starting with the effective weak Hamiltonian given by Eqs.~\eqref{eq:scalar Hamiltonian} and \eqref{eq:vector Hamiltonian}, one can get the helicity amplitudes of the decay. Since all types of the leptonic helicity amplitudes can be found in Ref.~\cite{Tanaka:2012nw}, we give only the hadronic helicity amplitudes. For the $(V-A)$-type current, we have~\cite{Gutsche:2015mxa}
\begin{equation}\label{eq:SM hadronic amplitudes}
H_{\lambda_2,\lambda_W}=H_{\lambda_2,\lambda_W}^V-H_{\lambda_2,\lambda_W}^A\,,\quad H_{\lambda_2,\lambda_W}^{V(A)}=\epsilon^{\dag\mu}(\lambda_W)\langle \Lambda_c,\lambda_2|\bar c\gamma_\mu(\gamma_\mu\gamma_5) b|\Lambda_b,\lambda_1\rangle\,,
\end{equation}
where $\lambda_2$ and $\lambda_W$ denote the helicities of the daughter baryon $\Lambda_c$ and the effective (axial-)vector-type current, respectively. The explicit expressions of $H_{\lambda_2,\lambda_W}$ in terms of the hadronic matrix elements defined by Eqs.~\eqref{eq:vector ffs} and \eqref{eq:axial-vector ffs} could be found in Ref.~\cite{Gutsche:2015mxa}. For the $(S-P)$-type current, the corresponding helicity amplitudes are given by~\cite{Shivashankara:2015cta}
\begin{align}\label{eq:scalar amplitudes}
H^{SP}_{\lambda_2,0}=&H^S_{\lambda_2,0}-H^P_{\lambda_2,0}\,,\\[0.2cm]
H^{SP}_{\pm\frac{1}{2},0}=&\frac{\sqrt{Q_+}}{m_b-m_c}\left(F_1^VM_-+F_3^V\frac{q^2}{M_1}\right)\pm\frac{\sqrt{Q_-}}
{m_b+m_c}
\left(F_1^AM_+-F_3^A\frac{q^2}{M_1}\right)\,,
\end{align}
where we use the abbreviations $M_\pm=M_{\Lambda_b}\pm M_{\Lambda_c}$ and $Q_\pm=M_{\pm}^2-q^2$. The hadronic helicity amplitudes of the (pseudo-)tensor-type current are defined as
\begin{equation}\label{eq:tensor definitions}
H_{\lambda_2,\lambda_W,\lambda_{W^\prime}}^{T}=\epsilon^{\dag\mu}(\lambda_W)\epsilon^{\dag\nu}
(\lambda_W^\prime)\langle \Lambda_c,\lambda_2|\bar c\,i\sigma_{\mu\nu}(1-\gamma_5)b|\Lambda_b,\lambda_1\rangle\,,
\end{equation}
and their explicit expressions, in terms of the hadronic matrix elements defined by Eqs.~\eqref{eq:tensor ffs} and \eqref{eq:pseudo-tensor ffs}, are given, respectively, by
\begin{align}\label{eq:tensor amplitudes}
H_{\frac{1}{2},+,0}^T=&-\sqrt{\frac{2}{q^2}}\left(f_T\sqrt{Q_+}M_-+g_T\sqrt{Q_-}M_+\right)\,,\nonumber\\[0.2cm]
H_{\frac{1}{2},+,-}^T=&-f_T\sqrt{Q_+}-g_T\sqrt{Q_-}\,,\nonumber\\[0.2cm]
H_{\frac{1}{2},+,t}^T=&-\sqrt{\frac{2}{q^2}}\left(f_T\sqrt{Q_-}M_++g_T\sqrt{Q_+}M_-\right)
+\sqrt{2q^2}\left(f_T^V\sqrt{Q_-}-g_T^V\sqrt{Q_+}\right)\,,\nonumber\\[0.2cm]
H_{\frac{1}{2},0,t}^T=&-f_T\sqrt{Q_-}-g_T\sqrt{Q_+}+f_T^V\sqrt{Q_-}M_+-g_T^V\sqrt{Q_+}M_-
+f_T^S\sqrt{Q_-}Q_++g_T^S\sqrt{Q_+}Q_-\,,\nonumber\\[0.2cm]
H_{-\frac{1}{2},+,-}^T=&f_T\sqrt{Q_+}-g_T\sqrt{Q_-}\,,\nonumber\\[0.2cm]
H_{-\frac{1}{2},0,-}^T=&\sqrt{\frac{2}{q^2}}\left(f_T\sqrt{Q_+}M_--g_T\sqrt{Q_-}M_+\right)\,,\nonumber\\[0.2cm]
H_{-\frac{1}{2},0,t}^T=&-f_T\sqrt{Q_-}+g_T\sqrt{Q_+}+f_T^V\sqrt{Q_-}M_+
+g_T^V\sqrt{Q_+}M_-+f_T^S\sqrt{Q_-}Q_+-g_T^S\sqrt{Q_+}Q_-\,,\nonumber\\[0.2cm]
H_{-\frac{1}{2},-,t}^T=&-\sqrt{\frac{2}{q^2}}\left(f_T\sqrt{Q_-}M_+-g_T\sqrt{Q_+}M_-\right)
+\sqrt{2q^2}\left(f_T^V\sqrt{Q_-}+g_T^V\sqrt{Q_+}\right)\,.
\end{align}
The helicity amplitudes satisfy the relations $H_{\lambda_2,\lambda_W,\lambda_W}^T=0$ and $H_{\lambda_2,\lambda_W,\lambda_{W^\prime}}^T=-H_{\lambda_2,\lambda_{W^\prime},\lambda_W}^T$, while all the others are found to be zero. Using the HQET relations given by Eq.~\eqref{eq:ffs relations}, we can further simplify these helicity amplitudes.

\subsection{Observables in $\Lambda_b\to\Lambda_c\ell\bar\nu_\ell$ decays}

Here we follow the conventions used in Refs.~\cite{Shivashankara:2015cta,Gutsche:2015mxa,Dutta:2015ueb}, and write the two-fold differential angular decay distribution as
\begin{equation}\label{eq:differential angular}
\frac{{\rm d}^2\Gamma(\Lambda_b\to\Lambda_c\ell\bar\nu_\ell)}{{\rm d}q^2\,{\rm d}\cos\theta_\ell}=N\left[A_1+\frac{m_\ell^2}{q^2}\left(A_{2}^V+A_{2}^T\right)+2A_3+
\frac{4m_\ell}{\sqrt{q^2}}A_4+A_5\right]\,,
\end{equation}
with
\begin{align}\label{eq:coefficients}
N=&\frac{G_F^2|V_{cb}|^2q^2|\vec{\mathbf p}_2|}{512\pi^3M_{\Lambda_b}^2}\left(1-\frac{m_\ell^2}{q^2}\right)^2\,,\nonumber\\[0.3cm]
A_1=&C_V^2\left[2\sin^2\theta_\ell\big(H_{\frac{1}{2},0}^2+H_{-\frac{1}{2},0}^2\big)+(1-\cos\theta_\ell)^2
H_{\frac{1}{2},+}^2+(1+\cos\theta_\ell)^2H_{-\frac{1}{2},-}^2\right]\,,\nonumber\\[0.3cm]
A_2^V=&C_V^2\Big[2\cos^2\theta_\ell\big(H_{\frac{1}{2},0}^2+H_{-\frac{1}{2},0}^2\big)+\sin^2\theta_\ell
\big(H_{\frac{1}{2},+}^2+H_{-\frac{1}{2},-}^2\big)+2\big(H_{\frac{1}{2},t}^2+H_{-\frac{1}{2},t}^2\big)\nonumber\\[0.1cm]
&\,\,\,\,-4\cos\theta_\ell \big(H_{\frac{1}{2},0}H_{\frac{1}{2},t}+H_{-\frac{1}{2},0} H_{-\frac{1}{2},t}\big)\Big] \,,\nonumber\\[0.3cm]
A_2^T=&\frac{C_T^2}{4}\Big[2\sin^2\theta_\ell\big(H_{\frac{1}{2},+,-}^{T2}+H_{\frac{1}{2},0,t}^{T2}+
H_{-\frac{1}{2},+,-}^{T2}+H_{-\frac{1}{2},0,t}^{T2}+2H_{\frac{1}{2},+,-}^TH_{\frac{1}{2},0,t}^T+
2H_{-\frac{1}{2},+,-}^TH_{-\frac{1}{2},0,t}^T\big)\nonumber\\[0.1cm]
&\,\,\,\,+(1+\cos\theta_\ell)^2\big(H_{-\frac{1}{2},0,-}^{T2}+H_{-\frac{1}{2},-,t}^{T2}+
2H_{-\frac{1}{2},0,-}^TH_{-\frac{1}{2},-,t}^T\big)\nonumber\\[0.1cm]
&\,\,\,\,+(1-\cos\theta_\ell)^2\big(H_{\frac{1}{2},+,0}^{T2}+H_{\frac{1}{2},+,t}^{T2}+
2H_{\frac{1}{2},+,0}^TH_{\frac{1}{2},+,t}^T\big)\Big]\,,\nonumber\\[0.3cm]
A_3=&\frac{C_T^2}{8}\Big[2\cos^2\theta_\ell\big(H_{\frac{1}{2},+,-}^{T2}\!+\!H_{\frac{1}{2},0,t}^{T2}
+H_{-\frac{1}{2},+,-}^{T2}\!+H_{-\frac{1}{2},0,t}^{T2}+
2H_{\frac{1}{2},+,-}^TH_{\frac{1}{2},0,t}^T+2H_{-\frac{1}{2},+,-}^TH_{-\frac{1}{2},0,t}^T\big)\nonumber\\[0.1cm]
&\,\,\,\,+\sin^2\theta_\ell\big(
H_{\frac{1}{2},+,0}^{T2}\!+\!H_{\frac{1}{2},+,t}^{T2}+H_{-\frac{1}{2},0,-}^{T2}+H_{-\frac{1}{2},-,t}^{T2}
+2H_{\frac{1}{2},+,0}^TH_{\frac{1}{2},+,t}^T+2H_{-\frac{1}{2},0,-}^TH_{-\frac{1}{2},-,t}^T\big)\Big]\nonumber\\[0.1cm]
+&C_S^2\Big(H^{SP2}_{\frac{1}{2},0}+H^{SP2}_{-\frac{1}{2},0}\Big)\,,\nonumber\\[0.3cm]
A_4=&C_VC_S\Big[-\cos\theta_\ell\big(H_{\frac{1}{2},0}H^{SP}_{\frac{1}{2},0}+H_{-\frac{1}{2},0}H^{SP}_{-\frac{1}{2},0}\big)
+\big(H_{\frac{1}{2},t}H^{SP}_{\frac{1}{2},0}+H_{-\frac{1}{2},t}H^{SP}_{-\frac{1}{2},0}\big)\Big]\nonumber\\[0.1cm]
+&C_VC_T\Big[\frac{\cos^2\theta}{2}\big(H_{\frac{1}{2},0}H^T_{\frac{1}{2},+,-}+
H_{\frac{1}{2},0}H^T_{\frac{1}{2},0,t}+H_{-\frac{1}{2},0}H^T_{-\frac{1}{2},+,-}+
H_{-\frac{1}{2},0}H^T_{-\frac{1}{2},0,t}\big)\nonumber\\[0.1cm]
&\hspace{0.9cm}-\frac{\cos\theta}{2}\big(H_{\frac{1}{2},t}H^T_{\frac{1}{2},+,-}+
H_{\frac{1}{2},t}H^T_{\frac{1}{2},0,t}+H_{-\frac{1}{2},t}H^T_{-\frac{1}{2},+,-}+
H_{-\frac{1}{2},t}H^T_{-\frac{1}{2},0,t}\big)\nonumber\\[0.1cm]
&\hspace{0.9cm}+\frac{(1-\cos\theta)^2}{4}\big(H_{\frac{1}{2},+}H^T_{\frac{1}{2},+,0}
+H_{\frac{1}{2},+}H^T_{\frac{1}{2},+,t}\big)\nonumber\\[0.1cm]
&\hspace{0.9cm}+\frac{(1+\cos\theta)^2}{4}\big(H_{-\frac{1}{2},-}H^T_{-\frac{1}{2},0,-}
+H_{-\frac{1}{2},-}H^T_{-\frac{1}{2},-,t}\big)\nonumber\\[0.1cm]
&\hspace{0.9cm}+\frac{\sin^2\theta}{4}\big(H_{\frac{1}{2},+}H^T_{\frac{1}{2},+,0}+
H_{\frac{1}{2},+}H^T_{\frac{1}{2},+,t}+H_{-\frac{1}{2},-}H^T_{-\frac{1}{2},0,-}+
H_{-\frac{1}{2},-}H^T_{-\frac{1}{2},-,t}\nonumber\\[0.1cm]
&\hspace{1.9cm}+2H_{\frac{1}{2},0}H^T_{\frac{1}{2},+,-}+2H_{\frac{1}{2},0}H^T_{\frac{1}{2},0,t}
+2H_{-\frac{1}{2},0}H^T_{-\frac{1}{2},+,-}+2H_{-\frac{1}{2},0}H^T_{-\frac{1}{2},0,t}\big)\Big]\,,\nonumber\\[0.3cm]
A_5=&-2C_SC_T\cos\theta\Big(H^{SP}_{\frac{1}{2},0}H^T_{\frac{1}{2},+,-}+H^{SP}_{\frac{1}{2},0}
H^T_{\frac{1}{2},0,t}+H^{SP}_{-\frac{1}{2},0}H^T_{-\frac{1}{2},+,-}
+H^{SP}_{-\frac{1}{2},0}H^T_{-\frac{1}{2},0,t}\Big)\,,
\end{align}
where $|\vec{\mathbf p}_2|=\sqrt{Q_{+}Q_{-}}/(2M_{\Lambda_b})$ is the $\Lambda_c$ momentum in the $\Lambda_b$ rest frame, $q^2$ the momentum transfer squared, and $\theta_\ell$ the polar angle of the lepton, as defined in Fig.~1 of Ref.~\cite{Gutsche:2015mxa}. Integrating out $\cos\theta_\ell$ in Eq.~\eqref{eq:differential angular}, one can then obtain the differential decay rate ${\rm d}\Gamma(\Lambda_b\to\Lambda_c\ell\bar\nu_{\ell})/{\rm d}q^2$. The above results are given for the scalar LQ scenario. For the vector LQ case, we need only to replace $C_V$ by $C_V^\prime$ given by Eq.~\eqref{eq:vector WC}, while setting $C_S$ and $C_T$ to zero.

With Eqs.~\eqref{eq:differential angular} and \eqref{eq:coefficients} at hand, we can get the following physical observables:
\begin{itemize}
  \item The differential and total branching fractions
  \begin{equation}
  \frac{{\rm d}\mathcal B(\Lambda_b\to\Lambda_c\ell\bar\nu_\ell)}{{\rm d}q^2}=\tau_{\Lambda_b}\frac{{\rm d}\Gamma(\Lambda_b\to\Lambda_c\ell\bar\nu_\ell)}{{\rm d}q^2}\,,\quad
  \mathcal B(\Lambda_b\to\Lambda_c\ell\bar\nu_\ell)=\int_{m_{\ell}^2}^{M_{-}^2} {\rm d}q^2 \frac{{\rm d}\mathcal B}{{\rm d}q^2}\,,
  \end{equation}
  where $\tau_{\Lambda_b}$ is the lifetime of $\Lambda_b$ baryon, and $m_{\ell}$ the lepton mass.
  \item The differential and integrated ratios
  \begin{equation}
  R_{\Lambda_c}(q^2)=\frac{{\rm d}\Gamma(\Lambda_b\to\Lambda_c\tau\bar\nu_\tau)/{\rm d}q^2}{{\rm d}\Gamma(\Lambda_b\to\Lambda_c\ell\bar\nu_\ell)/{\rm d}q^2}\,,\quad
  R_{\Lambda_c}=\frac{\int_{m_{\tau}^2}^{M_{-}^2} {\rm d}q^2{\rm d}\Gamma(\Lambda_b\to\Lambda_c\tau\bar\nu_\tau)/{\rm d}q^2}{\int_{m_{\ell}^2}^{M_{-}^2} {\rm d}q^2{\rm d}\Gamma(\Lambda_b\to\Lambda_c\ell\bar\nu_\ell)/{\rm d}q^2}\,.
  \end{equation}
  \item The lepton-side forward-backward asymmetry
  \begin{equation}
  A_{\rm FB}(q^2) = \frac{\int_{0}^{1} {\rm d}\cos\theta({\rm d}^2\Gamma/{\rm d}q^2{\rm d}\cos\theta)-\int_{-1}^{0}{\rm d}\cos\theta({\rm d}^2\Gamma/{\rm d}q^2{\rm d} \cos\theta )}{{\rm d}\Gamma/{\rm d}q^2}\,,
  \end{equation}
  defined as the relative difference between the differential decay rates where the angle $\theta_\ell$ is smaller or greater than $\pi/2$.
\end{itemize}
Once the individual helicity-dependent differential decay rates are calculated, which are collected in Appendix~\ref{sec:helicity rates}, we can obtain another two observables, the $q^2$-dependent longitudinal polarizations of $\Lambda_c$ baryon and $\tau$ lepton, which are defined, respectively, as
\begin{equation}
P_L^{\Lambda_c}(q^2)=\frac{{\rm d}\Gamma^{\lambda_2=1/2}/{\rm d}q^2-
    {\rm d}\Gamma^{\lambda_2=-1/2}/{\rm d}q^2}{{\rm d}\Gamma/{\rm d}q^2}\,,\quad P_L^{\tau}(q^2)=\frac{{\rm d}\Gamma^{\lambda_{\tau}=1/2}/{\rm d}q^2-
    {\rm d}\Gamma^{\lambda_{\tau}=-1/2}/{\rm d}q^2}{{\rm d}\Gamma/{\rm d}q^2}\,.
\end{equation}

Although the case with a polarized $\Lambda_b$ can bring in a number of new observables~\cite{Kadeer:2005aq,Bialas:1992ny}, we assume here that the parent baryon $\Lambda_b$ is unpolarized, based on the observation that the $\Lambda_b$ polarization in the LHCb setup is measured to be small and compatible with zero~\cite{Aaij:2013oxa}. This leaves with us the above interesting observables. Unlike the case for light leptons $\ell=e,\mu$, in which the phase space can be fully constrained~\cite{Aaij:2015bfa}, the decay $\Lambda_b\to\Lambda_c\tau\bar\nu_{\tau}$ poses several experimental challenges. Firstly, it is not possible to determine the $\Lambda_b$ momentum from the tagging side at the LHCb. Secondly, as the $\tau$ decays inside the detector, there exist at least two neutrinos in the final state, prohibiting a direct signal-side reconstruction. Finally, the fact that the $\Lambda_c$ is long lived and decays only weakly increases further the difficulty in determining the $\Lambda_b$ decay vertex. Thus, it is very challenging to determine the kinematic distributions on an event-by-event basis at the LHCb. However, the heuristic methods employed to determine approximately the $B$-meson momentum in $B\to D^\ast\tau\bar\nu_\tau$ decay~\cite{Aaij:2015yra} open the possibility to study the $b$-hadron decays with multiple missing particles in hadron colliders. Also, it might be feasible to measure this decay at the future $e^+e^-$ colliders, such as the ILC and CEPC, where two jets of hadrons are produced at the $Z^0$ peak and appears mostly in opposite side with a large boost. These facilities are also featured by the high reconstruction and tagging efficiencies, as well as the capability to measure the missing momentum.~\footnote{It should be mentioned that, even if measured, the $q^2$ distributions will be sculpted by the phase space cuts; for example, in practice one cannot integrate over all the helicity angle $\theta_{\ell}$, because minimum lepton-energy cuts might be needed to somewhat isolate the $\Lambda_c$ baryon and the lepton.}

As detailed recently by Ivanov, K\"{o}rner and Tranin in Ref.~\cite{Ivanov:2017mrj}, the information on the $\tau$ polarization can be extracted from the angular distribution of its subsequent decay modes, such as the hadronic $\tau\to\pi\nu_\tau$ and $\tau\to\rho\nu_\tau$ and the leptonic $\tau\to\mu\bar\nu_{\mu}\nu_{\tau}$ and $\tau\to e\bar\nu_{e}\nu_{\tau}$ decays. Especially, the analyzing power of the decay $\tau\to\pi\nu_\tau$ is found to be $100\%$~\cite{Ivanov:2017mrj,Bullock:1992yt}. It is particularly interesting to note that the Belle Collaboration has recently reported on the first measurement of the $\tau$ longitudinal polarization in the decay $B\to D^\ast\tau\bar\nu_\tau$ with the subsequent decays $\tau\to\pi\nu_\tau$ and $\tau\to\rho\nu_\tau$~\cite{Hirose:2016wfn}. Although the experimental environment is different, this pioneering measurement would be very beneficial for future detailed studies of the $\tau$ polarization at the LHC and future $e^+e^-$ colliders.

The $\Lambda_c$ polarization can also be probed by analyzing the angular decay distribution of its subsequent decays, among which the $\Lambda_c\to\Lambda\pi$ and $\Lambda_c\to\Lambda\ell\nu_{\ell}$ modes are of particular interest~\cite{Gutsche:2015mxa,Korner:1994nh,Gutsche:2015rrt}. As shown in Ref.~\cite{Gutsche:2015mxa}, the analyzing power of $\Lambda_c\to\Lambda\pi$ is close to maximal, and that of $\Lambda_c\to\Lambda\ell\nu_{\ell}$, although being not as large as the former, may still lead to reliable measurement. The biggest experimental challenge for the $\Lambda_c$ polarization measurement in the cascade decay $\Lambda_b\to\Lambda_c (\to\Lambda\pi)\ell\bar\nu_\ell$ is still how to reconstruct the $\Lambda_b$ rest frame mentioned already. But it is still hoped that, with more and more information on the $\Lambda_c$ decays from the Belle~\cite{Zupanc:2013iki} and BESIII~\cite{Ablikim:2015flg,Ablikim:2015prg,Ablikim:2016vqd} Collaborations, and sufficient statistics for the $\Lambda_b$ baryon at the LHC and future $e^+e^-$ colliders, the $\Lambda_c$ polarization could be measured in this decay.

\section{Numerical results and discussions}
\label{sec:numerical results}

\subsection{Input parameters}

In this section, we investigate the scalar and vector LQ effects on the aforementioned observables, to see if their effects are large enough to cause sizable deviations from the corresponding SM predictions. Firstly, we collect in Table~\ref{tab:inputs} all the input parameters used in this paper.

\begin{table}[t]
\setlength{\abovecaptionskip}{0pt}
\setlength{\belowcaptionskip}{10pt}
\begin{center}
\caption{\label{tab:inputs} Input parameters used in our numerical analyses.}
\renewcommand\arraystretch{1.3}
\tabcolsep 0.25in
\begin{tabular}{ccc}
\toprule
\toprule
Parameter & Value & Reference\\
\midrule
$G_F$ & $1.166378\times10^{-5}~{\rm GeV}^{-2}$ & \cite{Agashe:2014kda}\\
$\alpha_s(M_Z)$ & $0.1185\pm0.0006$ & \cite{Agashe:2014kda}\\
$M_Z$ & $91.188~{\rm GeV}$ & \cite{Agashe:2014kda}\\
$m_t$ & $(173.21\pm0.87)~{\rm GeV}$ & \cite{Agashe:2014kda}\\
$m_b(m_b)$ & $(4.18\pm0.03)~{\rm GeV}$ & \cite{Agashe:2014kda}\\
$m_c(m_c)$ & $(1.275\pm0.025)~{\rm GeV}$ & \cite{Agashe:2014kda}\\
$\tau_{\Lambda_b}$        & $1.466~{\rm ps}$ & \cite{Agashe:2014kda}\\
$M_{\Lambda_b}$ &   $5.61951~{\rm GeV}$ & \cite{Agashe:2014kda}\\
$M_{\Lambda_c}$ &   $2.28646~{\rm GeV}$ & \cite{Agashe:2014kda}\\
$m_{\tau}$ &   $1.7769~{\rm GeV}$ & \cite{Agashe:2014kda}\\
$m_{\mu}$ &   $105.66~{\rm MeV}$ & \cite{Agashe:2014kda}\\
$m_e$ &   $0.511~{\rm MeV}$ & \cite{Agashe:2014kda}\\
$|V_{cb}|$        & $(41.1\pm 1.3)\times10^{-3}$ & \cite{Agashe:2014kda}\\
\bottomrule
\bottomrule
\end{tabular}
\end{center}
\end{table}

\begin{table}[ht]
\setlength{\abovecaptionskip}{0pt}
\setlength{\belowcaptionskip}{10pt}
\begin{center}
\caption{\label{tab:form factors} Pole parameterizations of the $\Lambda_b\to\Lambda_c$ transition form factors for two values of $\kappa$ and for two choices of the continuum model in QCD sum rules~\cite{MarquesdeCarvalho:1999bqs}.}
\renewcommand\arraystretch{1.3}
\tabcolsep 0.25in
\begin{tabular}{cccc}
\toprule
\toprule
continuum model & $\kappa$ & $F_1^V(q^2)$  &  $-F_2^V(q^2)/M_{\Lambda_b}$\\
\midrule
rectangular & 1 & $6.66/(20.27-q^2)$  & $-0.21/(15.15-q^2)$ \\
rectangular & 2 & $8.13/(22.50-q^2)$  & $-0.22/(13.63-q^2)$\\
triangular  & 3 & $13.74/(26.68-q^2)$ & $-0.41/(18.65-q^2)$\\
triangular  & 4 & $16.17/(29.12-q^2)$ & $-0.45/(19.04-q^2)$\\
\bottomrule
\bottomrule
\end{tabular}
\end{center}
\end{table}

For the $\Lambda_b\to\Lambda_c$ transition form factors, we firstly use the results obtained in QCD sum rules~\cite{MarquesdeCarvalho:1999bqs}, together with the HQET relations among the form factors~\cite{Neubert:1993mb,Falk:1991nq,Mannel:1990vg}. Four types of form-factor parametrizations for two values of the parameter $\kappa$, which is introduced to account for deviations from the factorization
hypothesis for four-quark condensates, and for two choices of the continuum model are shown in Table~\ref{tab:form factors}. For a comparison, we also adopt the latest lattice QCD results for the (axial-)vector form factors~\cite{Detmold:2015aaa}, where the $q^2$ dependence of the form factors is parameterized in a simplified $z$ expansion~\cite{Bourrely:2008za}, modified to account for pion-mass and lattice-spacing dependence. All relevant formulae and input data can be found in Eq.~(79) and Tables~VII--IX of Ref.~\cite{Detmold:2015aaa}. While for the (pseudo-)tensor form factors, since lattice result is unavailable so far, we still use the HQET relations to relate them to the corresponding (axial-)vector ones.

\subsection{Numerical analyses}

We now give our predictions for the branching fractions $\mathcal B(\Lambda_b\to\Lambda_c\ell\bar\nu_\ell)$ and the ratio $R_{\Lambda_c}$ both within the SM and in the scalar and the vector  LQ scenarios in Table~\ref{tab:branching ratios} with the form factors taken from QCD sum rules~\cite{MarquesdeCarvalho:1999bqs}, and in Table~\ref{tab:lattice results} with the form factors taken from lattice QCD calculations~\cite{Detmold:2015aaa}. The theoretical uncertainties in Table~\ref{tab:branching ratios} come only from the NP parameters given by Eq.~\eqref{eq:parameter}, whereas in Table~\ref{tab:lattice results} we have also included the uncertainties from the form-factor parameters following the procedure recommended in~\cite{Detmold:2015aaa}. Specifically, we have taken into account the correlation matrices between the form-factor parameters, and calculate the central values, statistical uncertainties, and total systematic uncertainties of any observable depending on these parameters, according to Eqs.~(82)--(84) specified in Ref.~\cite{Detmold:2015aaa}.

\begin{table}[t]
\setlength{\abovecaptionskip}{0pt}
\setlength{\belowcaptionskip}{10pt}
\renewcommand\arraystretch{1.5}
\begin{center}
\tabcolsep 0.019in
\caption{\label{tab:branching ratios} Predictions for the branching fractions~(in unit of $10^{-2}$) and the ratio $R_{\Lambda_c}$ of $\Lambda_b\to\Lambda_c\ell\bar\nu_\ell\,(\ell=e/\tau)$ decays both within the SM and in the scalar/vector LQ scenarios, with the form factors taken from QCD sum rules~\cite{MarquesdeCarvalho:1999bqs}.}
\begin{tabular}[h]{c|c|c|c|c|c|c|c|c|c|c|c}
\hline
\hline
\multirow{3}*{$\kappa$}&\!\! $\mathcal B(\Lambda_b\to\Lambda_c e\bar\nu_e)$&  \multicolumn{5}{c|}{\!\!\!\!$\mathcal B(\Lambda_b\to\Lambda_c\tau\bar\nu_\tau)$} & \multicolumn{5}{c}{$R_{\Lambda_c}$}\\
\cline{2-12}
& \multirow{2}*{SM} & \multirow{2}*{SM} & \multicolumn{2}{c|}{scalar LQ} & \multicolumn{2}{c|}{vector LQ} & \multirow{2}*{SM} & \multicolumn{2}{c|}{scalar LQ} & \multicolumn{2}{c}{vector LQ}\\
\cline{4-7}\cline{9-12}
&&&$P_A$ & $P_C$ & $R_A$ & $R_B$ & & $P_A$ & $P_C$ & $R_A$ & $R_B$\\
\hline
1 & 2.50 & 0.74 & 0.95 & 0.93 & $0.95\pm0.05$ & $0.94\pm0.05$ & 0.30 & 0.38 & 0.37 & $0.38\pm0.02$ & $0.37\pm0.02$ \\
\hline
2 & 2.67 & 0.73 & 0.93 & 0.91 & $0.93\pm0.05$ & $0.92\pm0.05$ & 0.27 & 0.35 & 0.34 & $0.35\pm0.02$ & $0.34\pm0.02$ \\
\hline
3 & 5.16 & 1.39 & 1.77 & 1.73 & $1.78\pm0.09$ & $1.75\pm0.09$ & 0.27 & 0.35 & 0.34 & $0.35\pm0.02$ & $0.34\pm0.02$ \\
\hline
4 & 5.74 & 1.50 & 1.92 & 1.88 & $1.93\pm0.10$ & $1.90\pm0.10$ & 0.26 & 0.34 & 0.33 & $0.34\pm0.02$ & $0.33\pm0.02$ \\
\hline
\hline
\end{tabular}
\end{center}
\end{table}

\begin{table}[ht]
\setlength{\abovecaptionskip}{0pt}
\setlength{\belowcaptionskip}{10pt}
\renewcommand\arraystretch{1.5}
\begin{center}
\tabcolsep 0.12in
\caption{\label{tab:lattice results} Same as in Table~\ref{tab:branching ratios} but with the form factors taken from lattice QCD calculation~\cite{Detmold:2015aaa}.}
\begin{tabular}[h]{c|c|c|c}
\hline
\hline
$\mathcal B(\Lambda_b\to \Lambda_ce\bar\nu_e)$ & \multicolumn{2}{c|}{SM}& $5.34\pm0.33$\\
\hline
\multirow{5}*{$\mathcal B(\Lambda_b\to \Lambda_c\tau\bar\nu_\tau)$} & \multicolumn{2}{c|}{SM} & $1.77\pm0.09$\\
\cline{2-4}
&\multirow{2}*{scalar LQ}& $P_A$ & $2.26\pm 0.12$\\
\cline{3-4}
&& $P_C$ & $2.22\pm 0.12$\\
\cline{2-4}
&\multirow{2}*{vector LQ}& $R_A$ & $2.27\pm0.17$ \\
\cline{3-4}
&& $R_B$ &$2.24\pm0.17$ \\
\hline
\multirow{5}*{$R_{\Lambda_c}$} & \multicolumn{2}{c|}{SM} & $0.33\pm0.01$\\
\cline{2-4}
&\multirow{2}*{scalar LQ}& $P_A$ & $0.42\pm 0.01$\\
\cline{3-4}
&& $P_C$ & $0.42\pm 0.01$\\
\cline{2-4}
&\multirow{2}*{vector LQ}& $R_A$ & $0.43\pm0.02$\\
\cline{3-4}
&& $R_B$ & $0.42\pm0.02$\\
\hline
\hline
\end{tabular}
\end{center}
\end{table}

From the numerical results given in Tables~\ref{tab:branching ratios} and ~\ref{tab:lattice results}, we can draw the following conclusions:
\begin{itemize}
\item The branching fractions are very sensitive to the form-factor parameterizations used in QCD sum rules~\cite{MarquesdeCarvalho:1999bqs}. The triangular region (with $\kappa=3$ or $\kappa=4$) for the continuum model gives more reliable predictions compared to the rectangular one, because, within the SM, the former leads to consistent results with that obtained using the lattice-based form factors, and also with the experimental data $\mathcal B(\Lambda_b\to\Lambda_c e\bar\nu_e)=(6.5^{+3.2}_{-2.5})\%$~\cite{Agashe:2014kda}. Thus, from now on, we consider only the triangular continuum model with two different values of $\kappa$. The ratio $R_{\Lambda_c}$, on the other hand, is insensitive to the form-factor parameterizations, as is generally expected. It is also noted that the predicted $R_{\Lambda_c}$ using the lattice-based form factors is a little bit larger than that obtained from QCD sum rules, both within the SM and in the two LQ scenarios.

\item In the scalar LQ scenario, the predicted branching fraction $\mathcal B(\Lambda_b\to \Lambda_c\tau\bar\nu_\tau)$ is enhanced by about $28\%$ ($25\%$) in the $P_A$ ($P_C$) case, no matter the form factors are taken from QCD sum rules or from the lattice QCD calculations. The slight difference between the two solutions results from the $\sim1.2\%$ numerical difference in the dominated coefficient $|C_V^{\rm fit}|$, which has been discussed in Ref.~\cite{Li:2016vvp}. As the decay modes with light leptons ($\ell=e,\mu$) are assumed to be free from the scalar LQ contribution, the ratio $R_{\Lambda_c}$ is also enhanced by the corresponding percentages relative to the SM prediction.

\item In the vector LQ scenario, compared to the SM prediction, the branching fraction $\mathcal B(\Lambda_b\to \Lambda_c\tau\bar\nu_\tau)$ is found to be enhanced by about $28\%$ in the $R_A$ and by about $27\%$ in the $R_B$ case, respectively. To understand this, we should note that there is only one $(V-A)$ coupling in this scenario, and the resulting effective coefficients $C_V^\prime$, corresponding to the two solutions $R_A$ and $R_B$ (cf. Eq.~\eqref{eq:parameter}), are given, respectively, as
    \begin{equation}
     C_V^{\prime\,\text{fit}}=\left\{\begin{array}{cr}
                         \phantom{-}1.133\pm0.030, & \text{for $R_A$} \\
                                   -1.124\pm0.030, & \text{for $R_B$}
                                     \end{array}
     \right.\,.
     \end{equation}
     One can see clearly that, just like $C_V^{\rm fit}$ in scalar LQ scenario, $C_V^{\prime\,\rm fit}$ also has nearly the same absolute values for the two solutions $R_A$ and $R_B$, both enhancing the SM result by $\sim13\%$, but the sign of solution $R_B$ is flipped relative to the SM part.

\item As the two effective couplings $|C_V^{\rm fit}|$  and $|C_V^{\prime\,\rm fit}|$  are both enhanced by about $12\%\sim13\%$, compared to the SM part, they would give quite similar predictions for the other observables in $\Lambda_b\to\Lambda_c\tau\bar\nu_\tau$ decay.
\end{itemize}

\begin{figure}[t]
  \centering
  \subfigure[]{\includegraphics[width=3.0in]{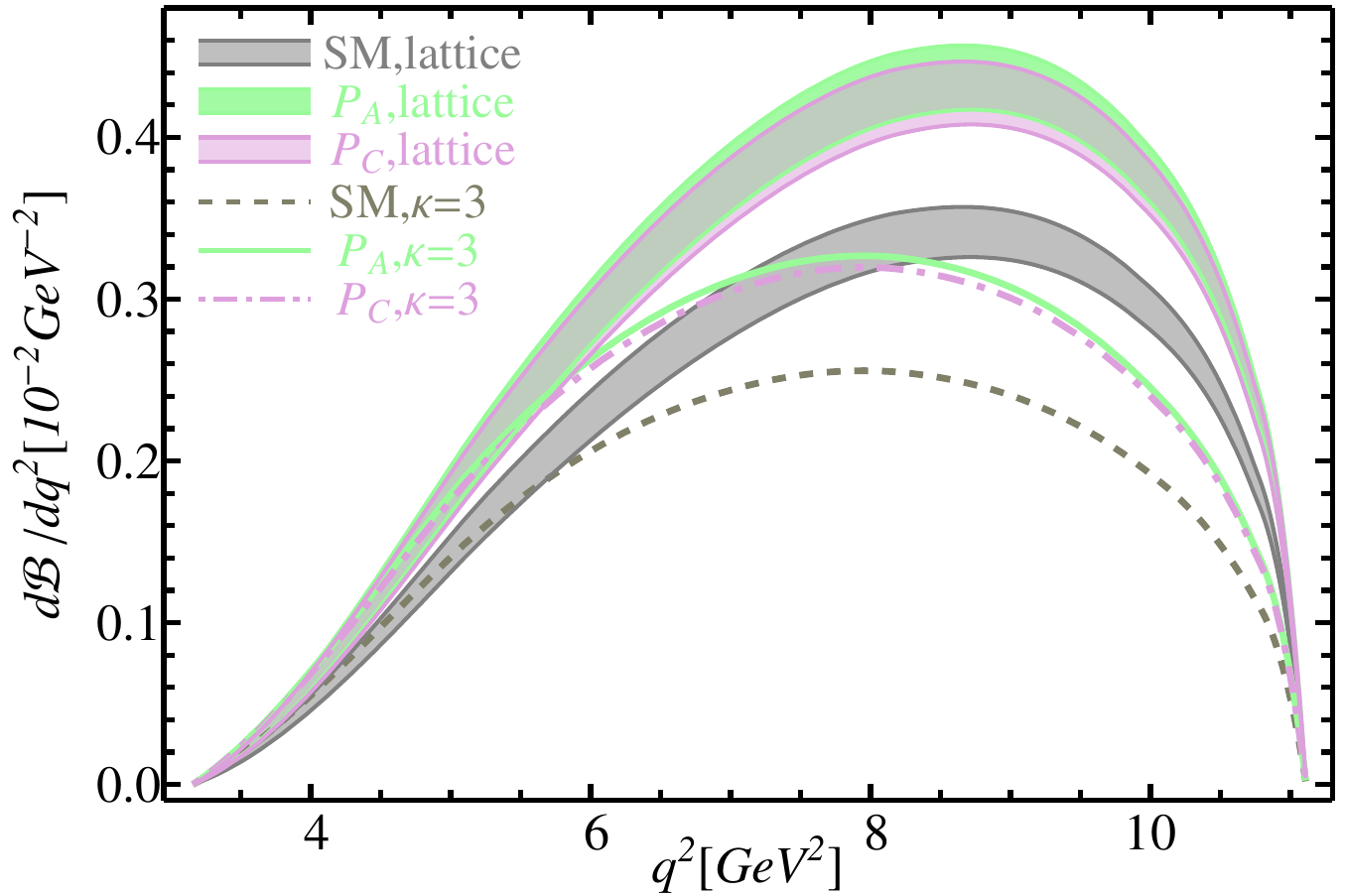}}
  \hspace{0.2in}
  \subfigure[]{\includegraphics[width=3.0in]{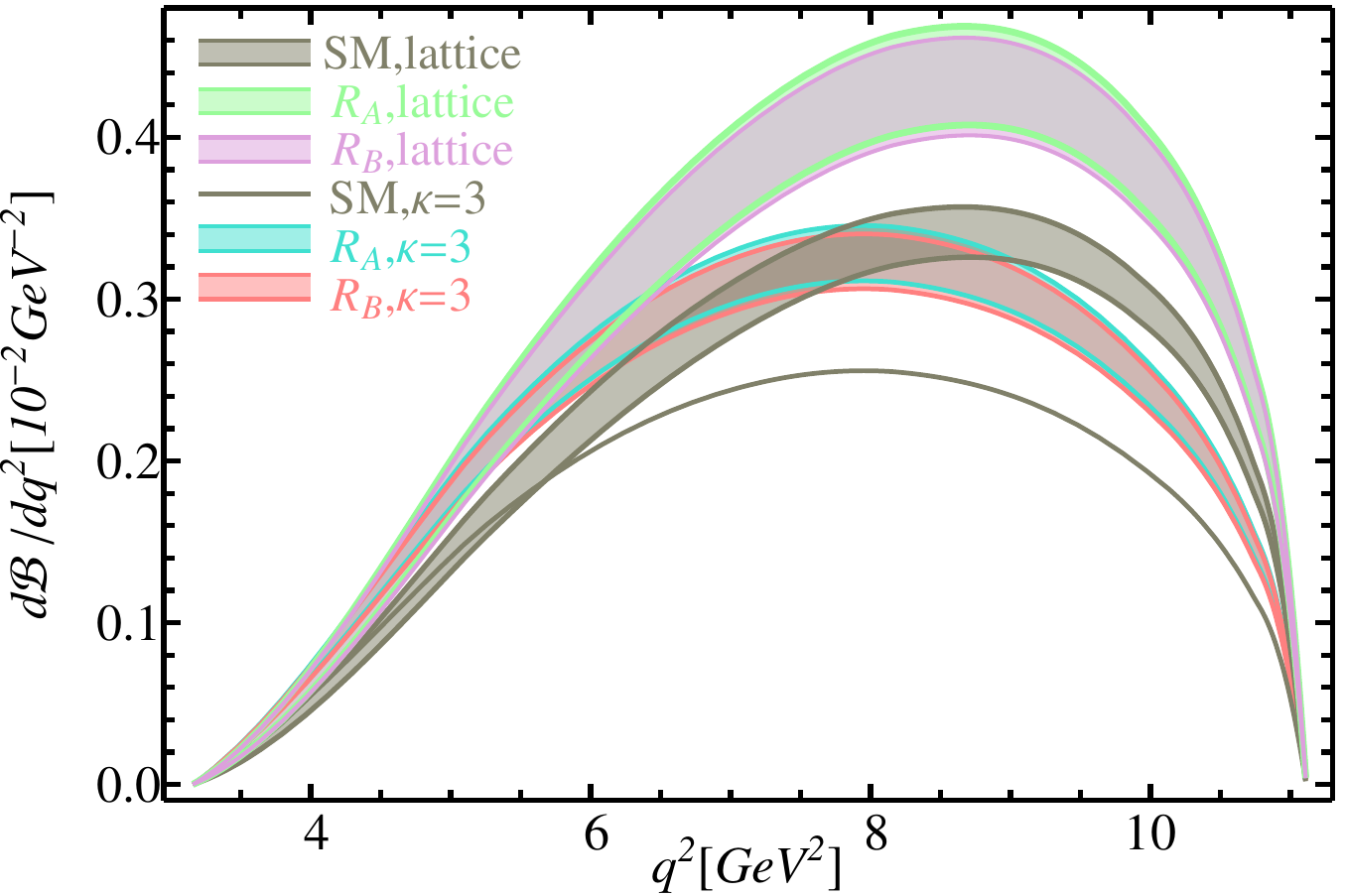}}\\[0.2cm]
  \subfigure[]{\includegraphics[width=3.0in]{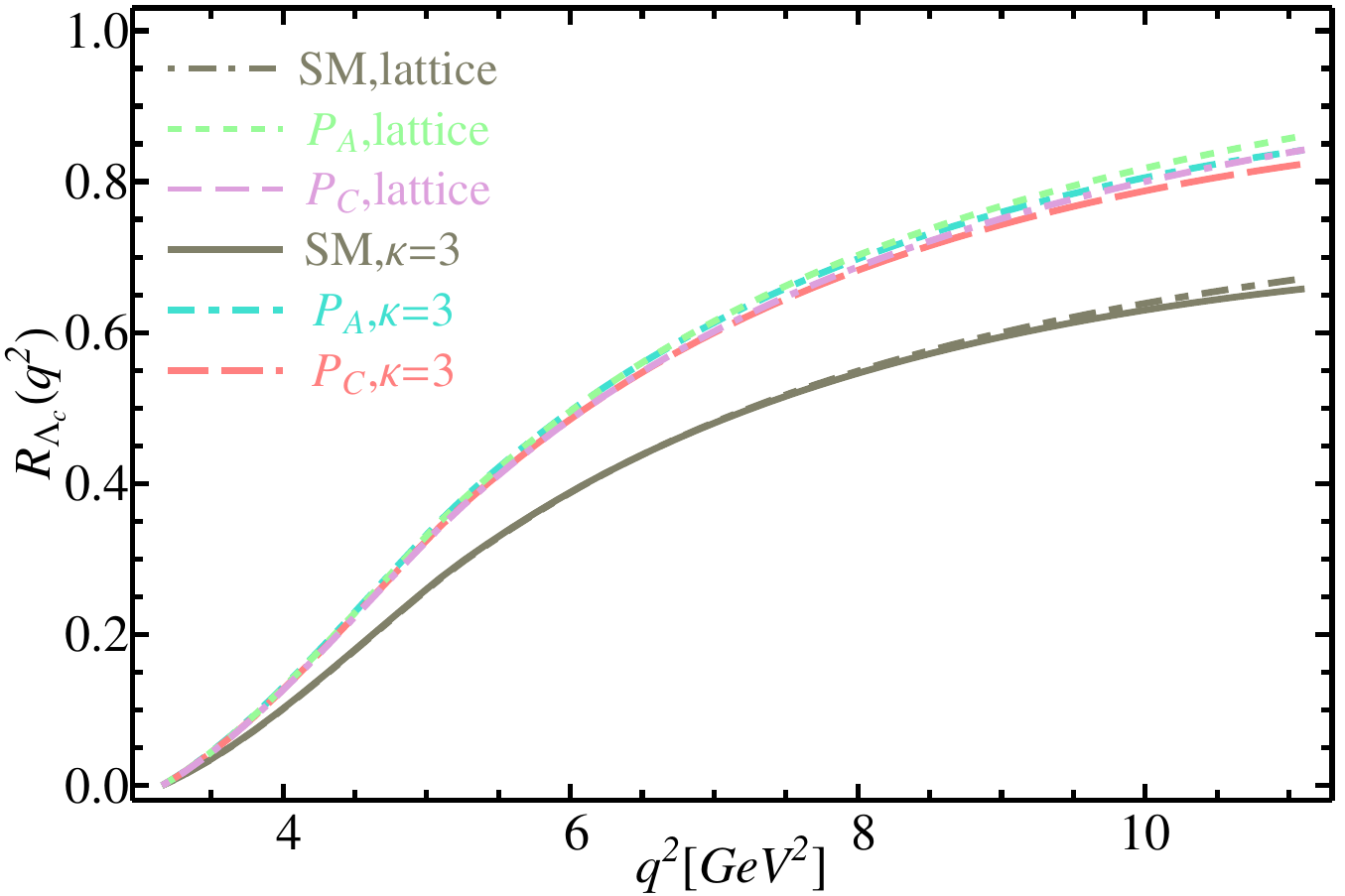}}
  \hspace{0.2in}
  \subfigure[]{\includegraphics[width=3.0in]{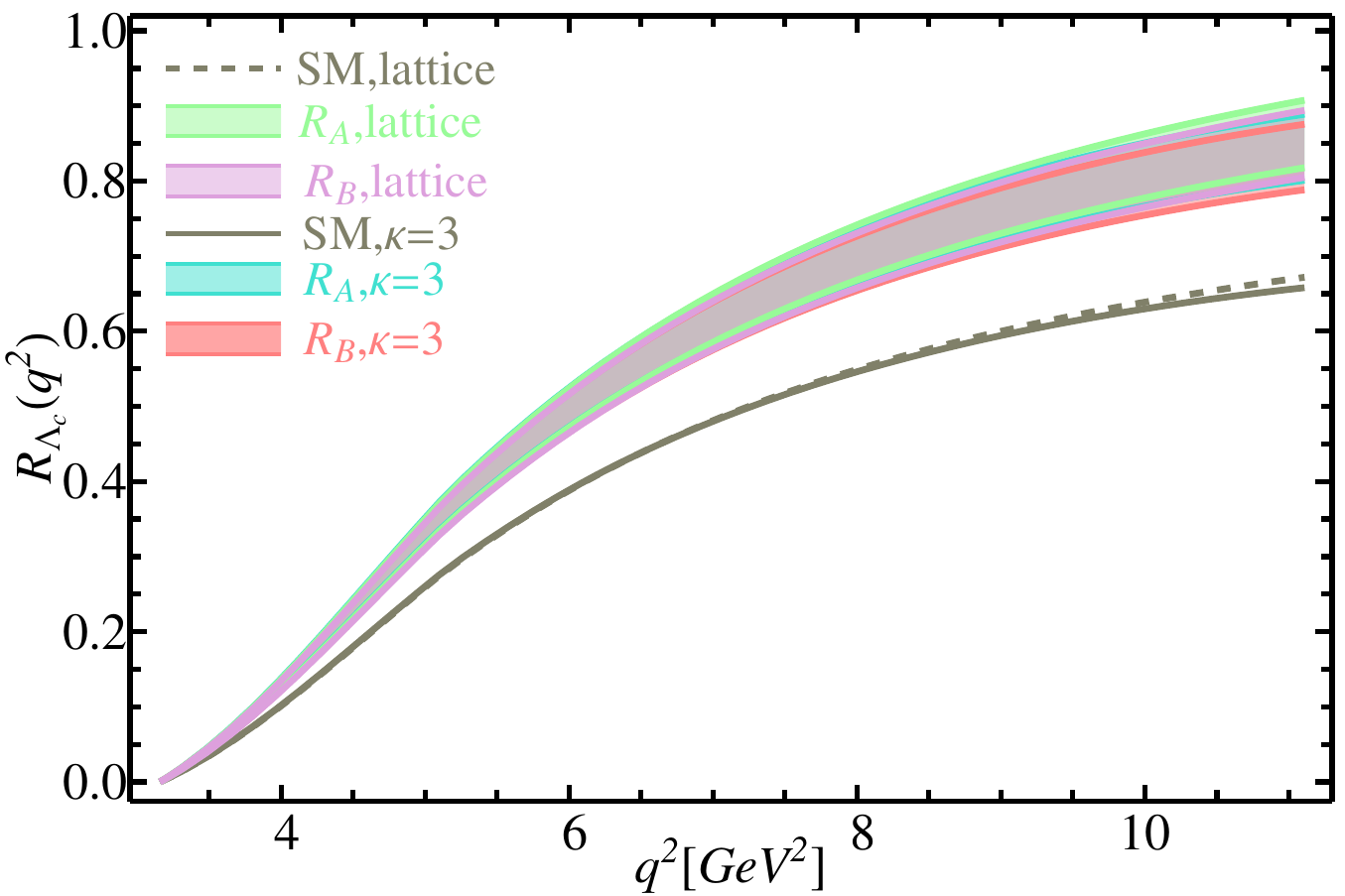}}
  \caption{\label{fig:dBr} The $q^2$ distributions of the differential branching fraction ${\rm d}\mathcal B(\Lambda_b\to\Lambda_c\tau\bar\nu_\tau)/{\rm d}q^2$ ((a): in the scalar and (b): in the vector scenario) and the ratio $R_{\Lambda_c}(q^2)$ ((c): in the scalar scenario and (d): in the vector scenario). The bands in (a) and (c) due to the uncertainties of form-factor parameters obtained in lattice QCD, and in (b) and (d) also include the varyings of the NP parameters in the vector scenario}
\end{figure}

The $q^2$ dependences of the differential branching fraction ${\rm d}\mathcal B(\Lambda_b\to\Lambda_c\tau\bar\nu_\tau)/{\rm d}q^2$ and the ratio $R_{\Lambda_c}(q^2)$ are displayed in Fig.~\ref{fig:dBr}, both within the SM and in the two LQ scenarios. As the results based on the form-factor parametrizations with $\kappa=3$ are similar to that with $\kappa=4$, we show only the case with $\kappa=3$. One can see that these two observables present the same features as the corresponding $q^2$-integrated ones discussed above: The predicted ${\rm d}\mathcal B(\Lambda_b\to\Lambda_c\tau\bar\nu_\tau)/{\rm d}q^2$ using the lattice-based form factors are a little bit larger than that based on QCD sum rules, and are enhanced at $q^2\sim9\,{\rm GeV}^2$ at most in both the two LQ scenarios. However, the ratio $R_{\Lambda_c}(q^2)$ is insensitive to the choices of the form factors.

\begin{figure}[t]
  \centering
  \subfigure[]{\includegraphics[width=2.16in]{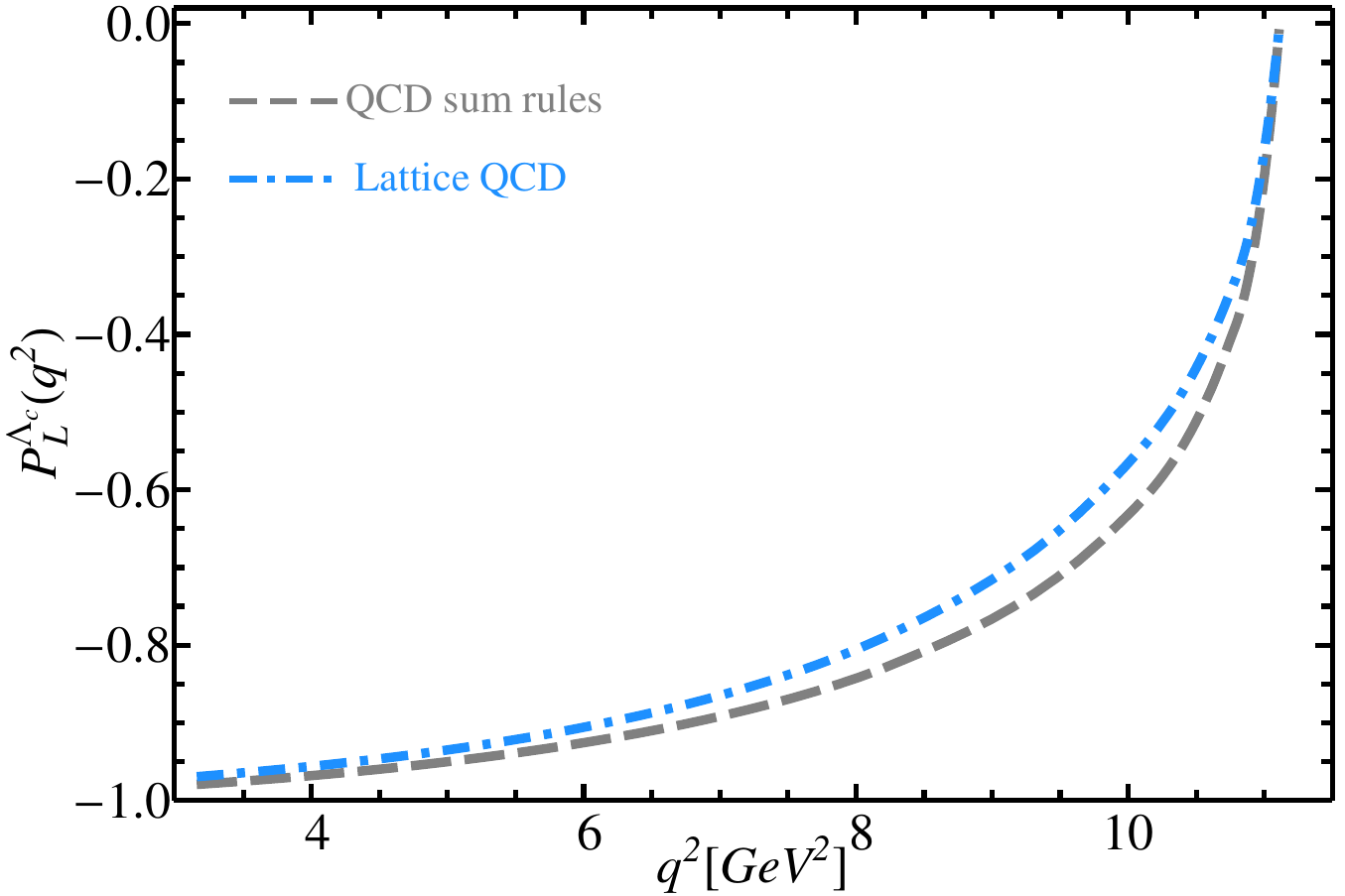}}\hspace{0.1cm}
  \subfigure[]{\includegraphics[width=2.16in]{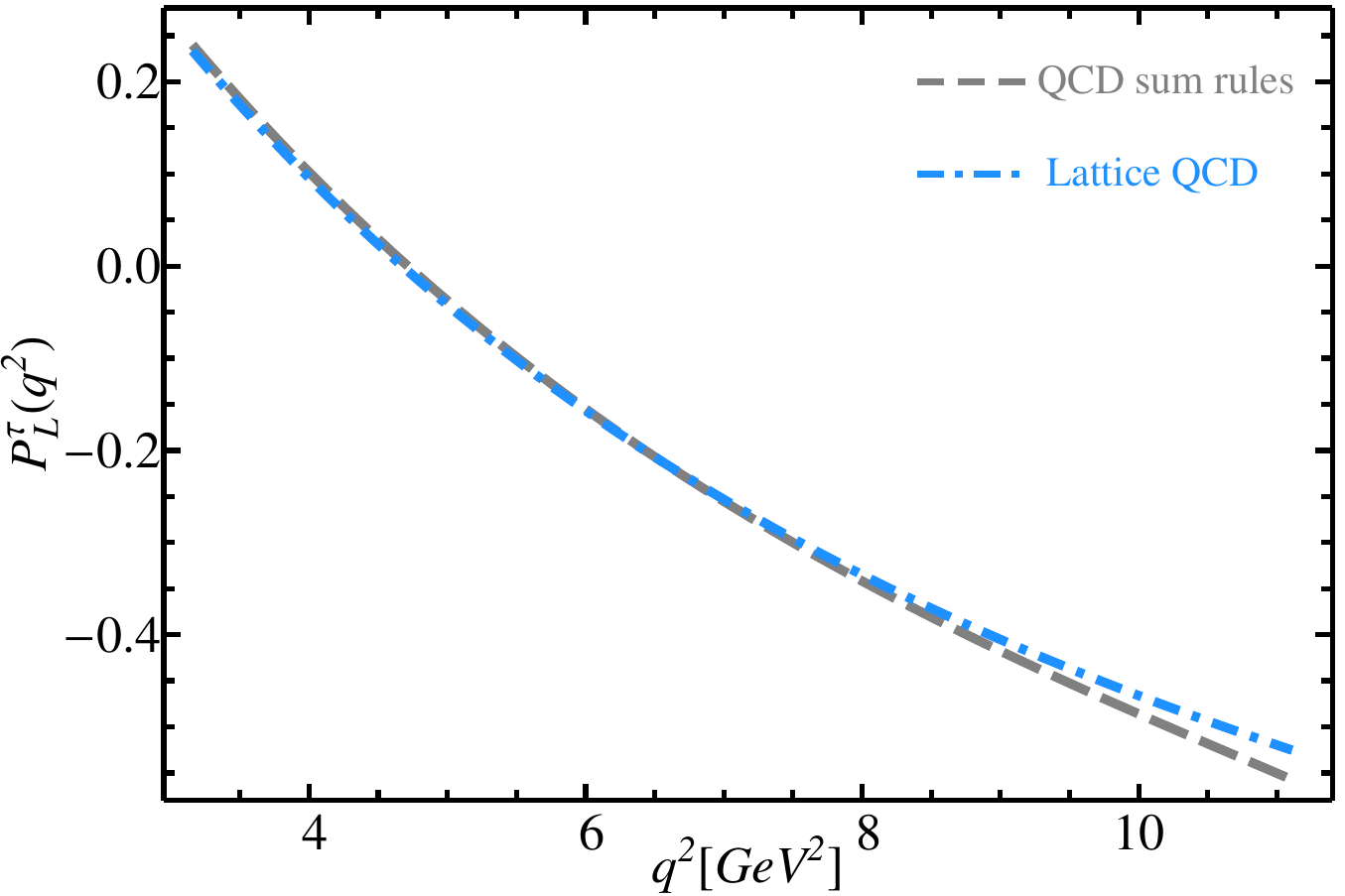}}\hspace{0.1cm}
  \subfigure[]{\includegraphics[width=2.16in]{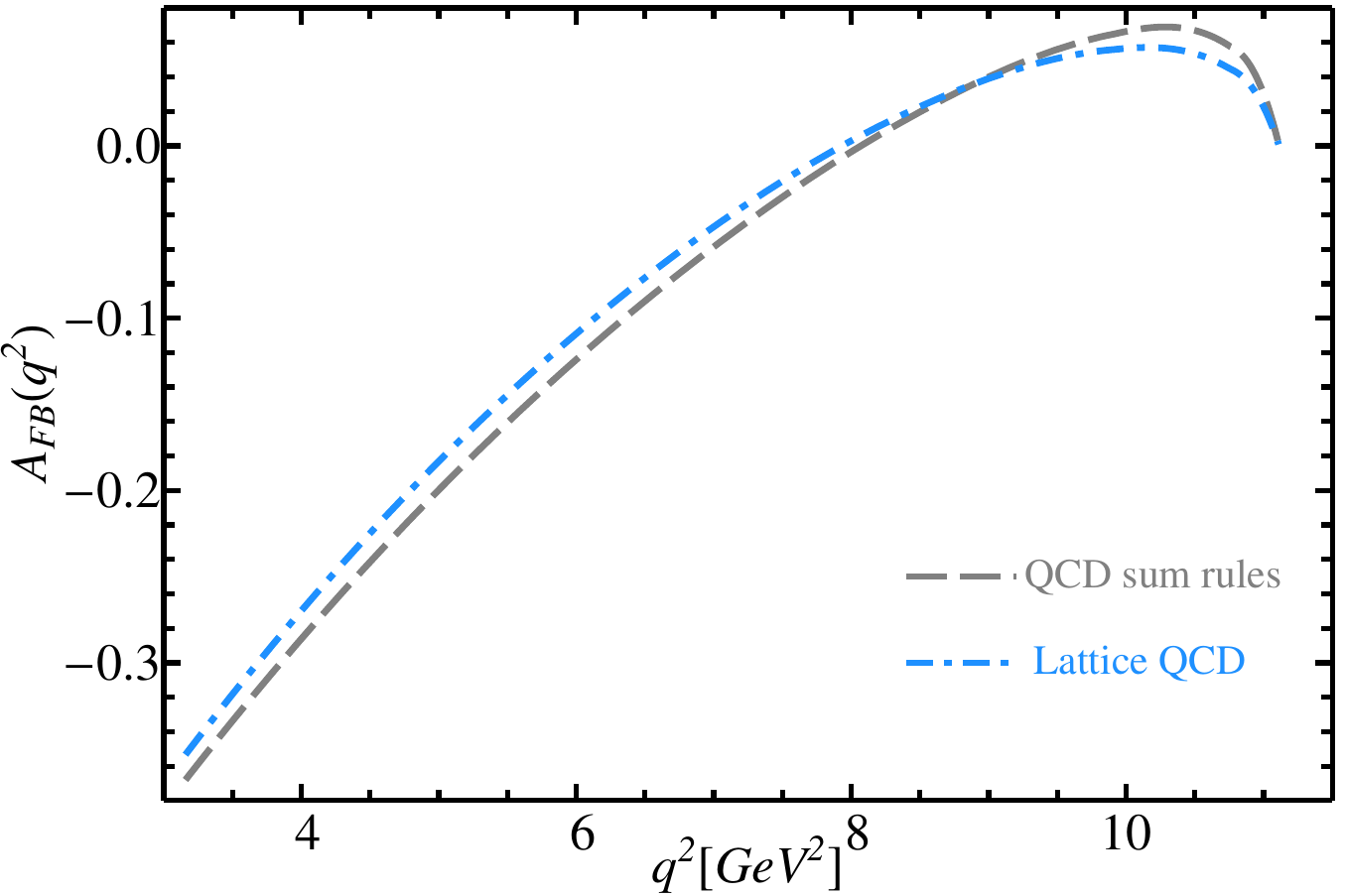}}
  \caption{\label{fig:polarizations} The $q^2$ dependences of the $\Lambda_c$ (a) and $\tau$ (b) longitudinal polarizations as well as the lepton-side forward-backward asymmetry (c), both within the SM and in the two LQ scenarios.}
\end{figure}

Finally, we show in Fig.~\ref{fig:polarizations} the $q^2$ dependences of the $\Lambda_c$ and $\tau$ longitudinal polarizations as well as the lepton-side forward-backward asymmetry. Since the resulting effective coefficients $C_V$ and $C_V^{\prime}$ of the dominated $(V-A)$ couplings appear both in the numerator and in the denominator of these ratios, the NP effects are cancelled exclusively. At the same time, the form-factor dependences of these observables are reduced to a large extent, and all the four cases in QCD sum rules ($\kappa=1,\cdots,4$) give almost the same curves for each observable, while being only slightly different from that obtained with the lattice-based form factors, as shown in Fig.~\ref{fig:polarizations}. As a consequence, all these three observables are insensitive to the two LQ scenarios and behave nearly the same as in the SM.

\section{Conclusions}\label{sec:conclusions}

As demonstrated in Refs.~\cite{Bauer:2015knc,Fajfer:2015ycq,Freytsis:2015qca}, both the scalar and vector LQ scenarios could explain the anomalies observed in $\bar B\to D^{(\ast)}\tau\bar\nu_\tau$ and $\bar B\to \bar K\ell^+\ell^-$ decays and, for each scenario, there exist two best-fit solutions for the operator coefficients, because the other two solutions for scalar LQ are already excluded by the $B_c^-\to\tau^-\bar\nu_\tau$ decay~\cite{Li:2016vvp}. To further explore these two interesting scenarios, in this paper, we have studies their effects in the semi-leptonic $\Lambda_b\to\Lambda_c\tau\bar\nu_\tau$ decay, which is induced by the same quark-level transition as in $\bar B\to D^{(\ast)}\tau\bar\nu_\tau$ decays. Besides the branching fraction $\mathcal B(\Lambda_b\to\Lambda_c\tau\bar\nu_\tau)$ and the ratio $R_{\Lambda_c}$, we have also discussed the $q^2$ distributions of these two observables, as well as the $\Lambda_c$ and $\tau$ longitudinal polarizations and the lepton-side forward-backward asymmetry, using the $\Lambda_b\to\Lambda_c$ transition form factors from both the QCD sum rules and the latest lattice QCD calculations. In addition, we have also discussed the feasibility for the measurements of these observables at the LHC and the future $e^+e^-$ colliders.

Using the best-fit solutions for the operator coefficients allowed by the current data of mesonic decays, we have found that the two LQ scenarios give the similar amounts of enhancements relative to the SM predictions for the branching fraction $\mathcal B(\Lambda_b\to\Lambda_c\tau\bar\nu_\tau)$ and the ratio $R_{\Lambda_c}$. The two best-fit solutions in each of the two scenarios are still found to be indistinguishable from each other based only on these two observables. On the other hand, both of the two LQ scenarios give nearly the same predictions as the SM for the $\Lambda_c$ and $\tau$ longitudinal polarizations, as well as the lepton-side forward-backward asymmetry. As a consequence, we conclude that, while the two LQ scenarios could be distinguished from the SM, it is quite difficult to distinguish between them using the semi-leptonic $\Lambda_b\to\Lambda_c\tau\bar\nu_\tau$ decay.

With large numbers of $\Lambda_b$ produced at the LHC and the future $e^+e^-$ colliders, we expect that the two LQ scenarios could be further tested, and even differentiated from other NP explanations to the $R_{D^{(\ast)}}$ anomalies, with the measurement of $\Lambda_b\to\Lambda_c\tau\bar\nu_\tau$ decay.

\section*{Acknowledgements}

We thank Prof. Yuehong Xie and Dr. Jiesheng Yu for useful discussions about the measurements of the $\Lambda_b\to\Lambda_c\tau\bar\nu_\tau$ decay at the LHC and future $e^+e^-$ colliders. The work is supported by the National Natural Science Foundation of China (NSFC) under contract Nos.~11675061, 11435003, 11225523 and 11521064. XL acknowledges the hospitality of the Munich Institute for Astro- and Particle Physics (MIAPP)
of the DFG cluster of excellence ``Origin and Structure of the Universe'', where this work was finalized.

\begin{appendix}

\section{$\Lambda_b \to \Lambda_c$ transition form factors}
\label{sec:form factors}

The hadronic matrix elements of the vector and axial-vector currents between the two spin-half baryons $\Lambda_b$ and $\Lambda_c$ can be parameterized in terms of three form factors, respectively, as~\cite{Gutsche:2015mxa}
\begin{align}
\langle \Lambda_c,\lambda_2|\bar c\gamma_\mu b|\Lambda_b,\lambda_1\rangle=&\bar u_2(p_2,\lambda_2)\left[F_1^V(q^2)\gamma_\mu-\frac{F_2^V(q^2)}{M_{\Lambda_b}}i\sigma_{\mu\nu}q^\nu+\frac{F_3^V(q^2)}
{M_{\Lambda_b}}q_\mu\right]u_1(p_1,\lambda_1)\,,\label{eq:vector ffs}\\[0.2cm]
\langle \Lambda_c,\lambda_2|\bar c\gamma_\mu\gamma_5 b|\Lambda_b,\lambda_1\rangle=&\bar u_2(p_2,\lambda_2)\left[F_1^A(q^2)\gamma_\mu-\frac{F_2^A(q^2)}{M_{\Lambda_b}}i\sigma_{\mu\nu}q^\nu+\frac{F_3^A(q^2)}
{M_{\Lambda_b}}q_\mu\right]\gamma_5u_1(p_1,\lambda_1)\,,\label{eq:axial-vector ffs}
\end{align}
where $\sigma_{\mu\nu}=\frac{i}{2}(\gamma_\mu\gamma_\nu-\gamma_\nu\gamma_\mu)$, $q=p_1-p_2$ is the four-momentum transfer, and $\lambda_i=\pm\frac{1}{2}\,(i=1,2)$ denote the helicities of the $\Lambda_b$ and $\Lambda_c$ baryons, respectively. Using the equations of motion, we can then obtain the hadronic matrix elements of the scalar and pseudo-scalar currents between the two baryons, which are given, respectively, as
\begin{align}
\langle \Lambda_c,\lambda_2|\bar cb|\Lambda_b,\lambda_1\rangle=&\frac{1}{m_b-m_c}\bar u_2(p_2,\lambda_2)\left[F_1^V(q^2)(M_1-M_2)+\frac{F_3^V(q^2)}
{M_{\Lambda_b}}q^2\right]u_1(p_1,\lambda_1)\,,\label{eq:scalar ffs}\\[0.2cm]
\langle \Lambda_c,\lambda_2|\bar c\gamma_5b|\Lambda_b,\lambda_1\rangle=&\frac{1}{m_b+m_c}\bar u_2(p_2,\lambda_2)\left[F_1^A(q^2)(M_1+M_2)-\frac{F_3^A(q^2)}
{M_{\Lambda_b}}q^2\right]\gamma_5u_1(p_1,\lambda_1)\,,\label{eq:pseudo-scalar ffs}
\end{align}
where $m_b$ and $m_c$ are the current quark masses evaluated at the scale $\mu\sim m_b$.

The hadronic matrix elements of the tensor and pseudo-tensor currents between the $\Lambda_b$ and $\Lambda_c$ baryons can be generally parameterized as~\cite{Chen:2001zc}
\begin{align}
\langle \Lambda_c,\lambda_2|\bar ci\sigma_{\mu\nu}b|\Lambda_b,\lambda_1\rangle=&\bar u_2(p_2,\lambda_2)
\left[f_T i\sigma_{\mu\nu} + f_T^V (\gamma_\mu q_\nu-\gamma_\nu q_\mu) + f_T^S (P_\mu q_\nu-P_\nu q_\mu)\right] u_1(p_1,\lambda_1)\,,\label{eq:tensor ffs}\\
\langle \Lambda_c,\lambda_2|\bar ci\sigma_{\mu\nu}\gamma_5b|\Lambda_b,\lambda_1\rangle=&\bar u_2(p_2,\lambda_2) \left[g_T i\sigma_{\mu\nu} + g_T^V (\gamma_\mu q_\nu-\gamma_\nu q_\mu) + g_T^S (P_\mu q_\nu-P_\nu q_\mu)\right] \gamma_5 u_1(p_1,\lambda_1)\,,\label{eq:pseudo-tensor ffs}
\end{align}
where $P=p_1+p_2$. The $\Lambda_b \to \Lambda_c$ transition form factors have also been studies based on the HQET~\cite{Neubert:1993mb,Falk:1991nq,Mannel:1990vg}, and the following relations among the form factors can be  found~\cite{Chen:2001zc}:
\begin{align}\label{eq:ffs relations}
&F_1^V=F_1^A=f_T=g_T\,,\\[0.2cm]
&F_2^V=F_2^A=-F_3^V=-F_3^A\,,\\[0.2cm]
&f_T^V\;=g_T^V=f_T^S=g_T^S=0\,.
\end{align}
An alternate helicity-based definition of the $\Lambda_b \to \Lambda_c$ form factors can be found in Ref.~\cite{Feldmann:2011xf}, and the explicit relations between these two sets of form factors are also given in Ref.~\cite{Feldmann:2011xf}.

In this paper, we use the results obtained both in the QCD sum rules~\cite{Feldmann:2011xf} and in the most recent lattice QCD calculation with $2 + 1$ dynamical flavours~\cite{Detmold:2015aaa}. However, since there are currently no lattice results for the (pseudo-)tensor form factors yet, we still use the following HQET relations to relate them to the corresponding (axial-)vector ones,
\begin{align}
f_T=g_T=F_1^V=\frac{(M_{\Lambda_b}+M_{\Lambda_c})^2f_+- q^2f_\perp}{(M_{\Lambda_b}+M_{\Lambda_c})^2-q^2}\,, \quad
f_T^V=g_T^V=f_T^S=g_T^S=0\,.
\end{align}

\section{Helicity-dependent differential decay rates}
\label{sec:helicity rates}

In order to discuss the $\Lambda_c$ and $\tau$ polarizations, we need the helicity-dependent differential decay rates, which are collected below (normalized by the prefactor $N$ defined in Eq.~\eqref{eq:coefficients}):
\begin{align}
\frac{{\rm d}\Gamma^{\lambda_2=1/2}}{{\rm d}q^2}=&\frac{m_\ell^2}{q^2}\Big[\frac{4}{3}C_V^2\big(H_{\frac{1}{2},+}^2+H_{\frac{1}{2},0}^2+
3H_{\frac{1}{2},t}^2\big)+\frac{2}{3}C_T^2\big(H^{T2}_{\frac{1}{2},+,-}+H^{T2}_{\frac{1}{2},0,t}
+H^{T2}_{\frac{1}{2},+,0}+H^{T2}_{\frac{1}{2},+,t}\nonumber\\[0.1cm]
&+2H^T_{\frac{1}{2},+,-}H^T_{\frac{1}{2},0,t}+2H^T_{\frac{1}{2},+,0}H^T_{\frac{1}{2},+,t}\big)\Big]
+\frac{8}{3}C_V^2\big(H_{\frac{1}{2},0}^2+H_{\frac{1}{2},+}^2\big)+4C_S^2H^{SP2}_{\frac{1}{2},0}\nonumber\\[0.1cm]
&+\frac{C_T^2}{3}\big(H^{T2}_{\frac{1}{2},+,-}+H^{T2}_{\frac{1}{2},0,t}+H^{T2}_{\frac{1}{2},+,0}
+H^{T2}_{\frac{1}{2},+,t}+2H^T_{\frac{1}{2},+,-}H^T_{\frac{1}{2},0,t}+2H^T_{\frac{1}{2},+,0}H^T_{\frac{1}{2},+,t}
\big)\nonumber\\[0.1cm]
+&\frac{4m_\ell}{\sqrt{q^2}}\Big[C_VC_T\big(H_{\frac{1}{2},0}H^T_{\frac{1}{2},+,-}
+H_{\frac{1}{2},0}H^T_{\frac{1}{2},0,t}+H_{\frac{1}{2},+}H^T_{\frac{1}{2},+,0}
+H_{\frac{1}{2},+}H^T_{\frac{1}{2},+,t}\big)\nonumber\\[0.1cm]
&+2C_VC_S\big(H_{\frac{1}{2},t}H^{SP}_{\frac{1}{2},0}\big)\Big]\,,\nonumber\\[0.2cm]
\frac{{\rm d}\Gamma^{\lambda_2=-1/2}}{{\rm d}q^2}=&\frac{m_\ell^2}{q^2}\Big[\frac{4}{3}C_V^2\big(H_{-\frac{1}{2},-}^2\!+\!H_{-\frac{1}{2},0}^2+
3H_{-\frac{1}{2},t}^2\big)\!+\!\frac{2}{3}C_T^2\big(H^{T2}_{-\frac{1}{2},+,-}\!+\!H^{T2}_{-\frac{1}{2},0,t}\!
+\!H^{T2}_{-\frac{1}{2},0,-}\!+\!H^{T2}_{-\frac{1}{2},-,t}\nonumber\\[0.1cm]
&+2H^T_{-\frac{1}{2},+,-}H^T_{-\frac{1}{2},0,t}+2H^T_{-\frac{1}{2},0,-}H^T_{-\frac{1}{2},-,t}\big)\Big]
+\frac{8}{3}C_V^2\big(H_{-\frac{1}{2},-}^2+H_{-\frac{1}{2},0}^2\big)+4C_S^2H^{SP2}_{-\frac{1}{2},0}\nonumber\\[0.1cm]
&+\frac{C_T^2}{3}\big(H^{T2}_{-\frac{1}{2},+,-}\!+\!H^{T2}_{-\frac{1}{2},0,t}\!+\!
H^{T2}_{-\frac{1}{2},0,-}\!+\!H^{T2}_{-\frac{1}{2},-,t}\!
+\!2H^T_{-\frac{1}{2},+,-}H^T_{-\frac{1}{2},0,t}\!+\!2H^T_{-\frac{1}{2},0,-}H^T_{-\frac{1}{2},-,t}\big)\nonumber\\[0.1cm]
+&\frac{4m_\ell}{\sqrt{q^2}}\Big[C_VC_T\big(H_{-\frac{1}{2},0}H^T_{-\frac{1}{2},+,-}
+H_{-\frac{1}{2},0}H^T_{-\frac{1}{2},0,t}+H_{-\frac{1}{2},-}H^T_{-\frac{1}{2},0,-}
+H_{-\frac{1}{2},-}H^T_{-\frac{1}{2},-,t}\big)\nonumber\\[0.1cm]
&+2C_VC_S\big(H_{-\frac{1}{2},t}H^{SP}_{-\frac{1}{2},0}\big)\Big]\,,\nonumber\\[0.2cm]
\frac{{\rm d}\Gamma^{\lambda_\tau=1/2}}{{\rm d}q^2}=&\frac{m_\ell^2}{q^2}C_V^2\Big[\frac{4}{3}\big(H_{\frac{1}{2},+}^2\!+\!H_{\frac{1}{2},0}^2\!+\!
H_{-\frac{1}{2},-}^2\!+\!H_{-\frac{1}{2},0}^2\big)\!+\!4\big(H_{\frac{1}{2},t}^2\!+\!H_{-\frac{1}{2},t}^2\big)\Big]
+4C_S^2\big(H^{SP2}_{\frac{1}{2},0}+H^{SP2}_{-\frac{1}{2},0}\big)\nonumber\\[0.1cm]
&+\frac{C_T^2}{3}\Big[H^{T2}_{\frac{1}{2},+,-}\!+\!H^{T2}_{\frac{1}{2},0,t}+H^{T2}_{\frac{1}{2},+,0}+
H^{T2}_{\frac{1}{2},+,t}+H^{T2}_{-\frac{1}{2},+,-}+H^{T2}_{-\frac{1}{2},0,t}+H^{T2}_{-\frac{1}{2},0,-}\!+
H^{T2}_{-\frac{1}{2},-,t}\nonumber\\[0.1cm]
&+2\big(H^T_{\frac{1}{2},+,-}H^T_{\frac{1}{2},0,t}+H^T_{\frac{1}{2},+,0}H^T_{\frac{1}{2},+,t}+
H^T_{-\frac{1}{2},+,-}H^T_{-\frac{1}{2},0,t}+H^T_{-\frac{1}{2},0,-}H^T_{-\frac{1}{2},-,t}\big)\Big]\nonumber\\[0.1cm]
+&\frac{4m_\ell}{3\sqrt{q^2}}\Big[6C_VC_S\big(H_{\frac{1}{2},t}H^{SP}_{\frac{1}{2},0}+H_{-\frac{1}{2},t}
H^{SP}_{-\frac{1}{2},0}\big)+C_VC_T\big(H_{\frac{1}{2},0}H^T_{\frac{1}{2},+,-}+
H_{\frac{1}{2},0}H^T_{\frac{1}{2},0,t}\nonumber\\[0.1cm]
&+H_{\frac{1}{2},+}H^T_{\frac{1}{2},+,0}+H_{\frac{1}{2},+}H^T_{\frac{1}{2},+,t}+
H_{-\frac{1}{2},0}H^T_{-\frac{1}{2},+,-}+H_{-\frac{1}{2},0}H^T_{-\frac{1}{2},0,t}+
H_{-\frac{1}{2},-}H^T_{-\frac{1}{2},0,-}\nonumber\\[0.1cm]
&+H_{-\frac{1}{2},-}H^T_{-\frac{1}{2},-,t}\big)\Big]\,,\nonumber\\[0.2cm]
\frac{{\rm d}\Gamma^{\lambda_\tau=-1/2}}{{\rm d}q^2}=&\frac{8C_V^2}{3}\big(H_{\frac{1}{2},+}^2\!+\!H_{\frac{1}{2},0}^2\!+\!H_{-\frac{1}{2},-}^2\!+
\!H_{-\frac{1}{2},0}^2\big)+\!\frac{2m_\ell^2}{3q^2}C_T^2\Big[H^{T2}_{\frac{1}{2},+,-}\!+\!H^{T2}_{\frac{1}{2},0,t}\!+
\!H^{T2}_{\frac{1}{2},+,0}\!+\!H^{T2}_{\frac{1}{2},+,t}\!\nonumber\\[0.1cm]
&+\!H^{T2}_{-\frac{1}{2},+,-}\!+\!H^{T2}_{-\frac{1}{2},0,t}
+H^{T2}_{-\frac{1}{2},0,-}\!+\!H^{T2}_{-\frac{1}{2},-,t}+2\big(H^T_{\frac{1}{2},+,-}H^T_{\frac{1}{2},0,t}
+H^T_{\frac{1}{2},+,0}H^T_{\frac{1}{2},+,t}\nonumber\\[0.1cm]
&+H^T_{-\frac{1}{2},+,-}H^T_{-\frac{1}{2},0,t}+H^T_{-\frac{1}{2},0,-}
H^T_{-\frac{1}{2},-,t}\big)\Big]\nonumber\\[0.1cm]
+&\frac{8m_\ell}{3\sqrt{q^2}}C_VC_T\big(H_{\frac{1}{2},0}H^T_{\frac{1}{2},+,-}+
H_{\frac{1}{2},0}H^T_{\frac{1}{2},0,t}+H_{\frac{1}{2},+}H^T_{\frac{1}{2},+,0}+
H_{\frac{1}{2},+}H^T_{\frac{1}{2},+,t}\nonumber\\[0.1cm]
&+H_{-\frac{1}{2},0}H^T_{-\frac{1}{2},+,-}+H_{-\frac{1}{2},0}H^T_{-\frac{1}{2},0,t}+
H_{-\frac{1}{2},-}H^T_{-\frac{1}{2},0,-}+H_{-\frac{1}{2},-}H^T_{-\frac{1}{2},-,t}\big)\,.
\end{align}

\end{appendix}



\end{document}